\documentclass[preprint]{ptephy_v1}

\preprintnumber{XXXX-XXXX} 

\usepackage{cancel}

\begin{document}

\title{The cross-section measurement for the $^3{\rm H}(e,e'K^+)nn\Lambda$ reaction}

\newcommand*{\Kyoto}{Graduate School of Science, Kyoto University, Kyoto 606-8502 Japan}
\newcommand*{\Tohoku}{Graduate School of Science, Tohoku University, Sendai, Miyagi 980-8578, Japan}
\newcommand*{\Hampton}{Department of Physics, Hampton University, Virginia 23668, USA}
\newcommand*{\JLab}{Thomas Jefferson National Accelerator Facility, Newport News, Virginia 23606, USA}
\newcommand*{\Virginia}{Department of Physics, University of Virginia, Charlottesville, Virginia 22904, USA}
\newcommand*{\Flinders}{School of Chemical and Physical Science, Flinders University, GPO Box 2100, Adelaide 5001, Australia}
\newcommand*{\Zagreb}{Department of Physics \& Department of Applied Physics, University of Zagreb, HR-10000 Zagreb, Croatia}
\newcommand*{\California}{Physics and Astronomy Department, California State University, Los Angeles, Califonia 90032, USA}
\newcommand*{\William}{Department of Physics, The College of William and Mary, Virginia 23185, USA}
\newcommand*{\Tennessee}{Department of Physics, University of Tennessee, Knoxville, Tennessee 37996, USA}
\newcommand*{\INFN}{INFN Sezione di Catania, Catania 95123, Italy}
\newcommand*{\Mississippi}{Department of Physics, Mississippi State University, Mississippi State, Mississippi 39762, USA }
\newcommand*{\Florida}{Department of Physics, Florida International University, Miami, Florida 33199, USA}
\newcommand*{\MichiganDepart}{Department of Physics and Astronomy, Michigan State University, East Lansing, Michigan 48824, USA}
\newcommand*{\MichiganLab}{National Superconducting Cyclotron Laboratory, Michigan State University, East Lansing, MI 48824, USA}
\newcommand*{\MIT}{Department of Physics, Massachusetts Institute of Technology, Cambridge, Massachusetts 02139, USA}
\newcommand*{\NewYork}{Department of Physics, State University of New York, Stony Brook, New York 11794, USA}
\newcommand*{\Connecticut}{Department of Physics, University of Connecticut, Storrs, Connecticut 06269, USA}
\newcommand*{\LosAlamos}{Theoretical Division, Los Alamos National Laboratory, Los Alamos, New Mexico 87545, USA}
\newcommand*{\Sezione}{INFN, Sezione di Roma, 00185, Rome, Italy}
\newcommand*{\Sapienza}{INFN, Sezione di Roma and Sapienza University of Rome, 00185, Rome, Italy}
\newcommand*{\SapienzaUniv}{Sapienza University of Rome, I-00185, Rome, Italy}
\newcommand*{\Superiore}{Istituto Superiore di Sanit\`{a}, 00161, Rome, Italy}
\newcommand*{\Kent}{Department of Physics, Kent State University, Kent, Ohio 44242, USA}
\newcommand*{\Old}{Department of Physics, Old Dominion University, Norfolk, Virginia 23529, USA}
\newcommand*{\NewHampshire}{Department of Physics, University of New Hampshire, Durham, New Hampshire 03824, USA}
\newcommand*{\BNL}{Nuclear Science Division, Lawrence Berkeley National Laboratory, Berkeley, CA 94720, USA}
\newcommand*{\Columbia}{Department of Physics, Columbia University, New York, New York 10027, USA}
\newcommand*{\Manitoba}{Department of Physics and Astronomy, University of Manitoba, Winnipeg, Manitoba R3T 2N2, Canada}
\newcommand*{\Ljubljana}{Faculty of Mathematics and Physics, University of Ljubljana, 1000 Ljubljana, Slovenia}
\newcommand*{\Jozef}{Jo\v{z}ef Stefan Institute, Ljubljana, Slovenia}
\newcommand*{\Johannes}{Institut f\"{u}r Kernphysik, Johannes Gutenberg-Universit\"{a}t Mainz, DE-55128 Mainz, Germany}
\newcommand*{\Christopher}{Department of Physics, Computer Science and Engineering, Christopher Newport University, Newport News, Virginia 23606, USA}
\newcommand*{\Hue}{University of Education, Hue University, Hue City, Vietnam}
\newcommand*{\Argonne}{Physics Division, Argonne National Laboratory, Lemont, Illinois 60439, USA}
\newcommand*{\Metropolitana}{Escuela de Ciencias Tecnologa, Universidat Metropolitana, San Juan 00928, Puerto Rico}
\newcommand*{\PuertoRico}{Divisi\'{o}n de Ciencias y Tecnologia, Universidad Ana G. M\'{e}ndez, Recinto de Cupey, San Juan 00926, Puerto Rico}
\newcommand*{\VirginiaMilitary}{Department of Physics \& Astronomy, Virginia Military Institute, Lexington, Virginia 24450, USA}
\newcommand*{\Syracuse}{Department of Physics, Syracuse University, New York, New York 10016, USA}
\newcommand*{\Glasgow}{School of Physics \& Astronomy, University of Glasgow, Glasgow, G12 8QQ, Scotland, UK}

\author[1]{K.~N.~Suzuki \thanks{Email: suzuki.kazuki.83v@st.kyoto-u.ac.jp}}
\author[1,2]{T.~Gogami \thanks{Email: gogami@jlab.org}}
\author[3]{B.~Pandey}
\author[2]{K.~Itabashi}
\author[2]{S.~Nagao}
\author[2]{K.~Okuyama}
\author[2]{S.~N.~Nakamura}
\author[3,4]{L.~Tang}
\author[5]{D.~Abrams}
\author[2]{T.~Akiyama}
\author[6]{D.~Androic}
\author[7]{K.~Aniol}
\author[8]{C.~Ayerbe~Gayoso}
\author[9]{J.~Bane}
\author[8]{S.~Barcus}
\author[9]{J.~Barrow}
\author[10]{V.~Bellini}
\author[11]{H.~Bhatt}
\author[11]{D.~Bhetuwal}
\author[3]{D.~Biswas}
\author[4]{A.~Camsonne}
\author[12]{J.~Castellanos}
\author[4]{J-P.~Chen}
\author[8]{J.~Chen}
\author[4]{S.~Covrig}
\author[13,14]{D.~Chrisman}
\author[15]{R.~Cruz-Torres}
\author[16]{R.~Das}
\author[17]{E.~Fuchey}
\author[5]{K.~Gnanvo}
\author[10,18]{F.~Garibaldi}
\author[3]{T.~Gautam}
\author[4]{J.~Gomez}
\author[3]{P.~Gueye}
\author[19]{T.~J.~Hague}
\author[4]{O.~Hansen}
\author[4]{W.~Henry}
\author[20]{F.~Hauenstein}
\author[4]{D.~W.~Higinbotham}
\author[20]{C.~E.~Hyde}
\author[2]{M.~Kaneta}
\author[4]{C.~Keppel}
\author[16]{T.~Kutz}
\author[3]{N.~Lashley-Colthirst}
\author[21,22]{S.~Li}
\author[23]{H.~Liu}
\author[24]{J.~Mammei}
\author[12]{P.~Markowitz}
\author[4]{R.~E.~McClellan}
\author[10,25]{F.~Meddi}
\author[4]{D.~Meekins}
\author[4]{R.~Michaels}
\author[26,27,28]{M.~Mihovilovi\textrm{\v{c}}}
\author[29]{A.~Moyer}
\author[15,30]{D.~Nguyen}
\author[19]{M.~Nycz}
\author[8]{V.~Owen}
\author[5]{C.~Palatchi}
\author[16]{S.~Park}
\author[6]{T.~Petkovic}
\author[5]{S.~Premathilake}
\author[31]{P.~E.~Reimer}
\author[12]{J.~Reinhold}
\author[31]{S.~Riordan}
\author[32]{V.~Rodriguez}
\author[33]{C.~Samanta}
\author[21]{S.~N.~Santiesteban}
\author[4]{B.~Sawatzky}
\author[26,27]{S.~\textrm{{\v{S}}}irca}
\author[21]{K.~Slifer}
\author[19]{T.~Su}
\author[34]{Y.~Tian}
\author[2]{Y.~Toyama}
\author[2]{K.~Uehara}
\author[10]{G.~M.~Urciuoli}
\author[13,14]{D.~Votaw}
\author[35]{J.~Williamson}
\author[4]{B.~Wojtsekhowski}
\author[4]{S.~A.~Wood}
\author[21]{B.~Yale}
\author[31]{Z.~Ye}
\author[5]{J.~Zhang}
\author[5]{X.~Zheng}

\affil[1]{\Kyoto}
\affil[2]{\Tohoku}
\affil[3]{\Hampton}
\affil[4]{\JLab}
\affil[5]{\Virginia}
\affil[6]{\Zagreb}
\affil[7]{\California}
\affil[8]{\William}
\affil[9]{\Tennessee}
\affil[10]{\Sezione}
\affil[11]{\Mississippi}
\affil[12]{\Florida}
\affil[13]{\MichiganDepart}
\affil[14]{\MichiganLab}
\affil[15]{\MIT}
\affil[16]{\NewYork}
\affil[17]{\Connecticut}
\affil[18]{\Superiore}
\affil[19]{\Kent}
\affil[20]{\Old}
\affil[21]{\NewHampshire}
\affil[22]{\BNL}
\affil[23]{\Columbia}
\affil[24]{\Manitoba}
\affil[25]{\SapienzaUniv} 
\affil[26]{\Ljubljana}
\affil[27]{\Jozef}
\affil[28]{\Johannes}
\affil[29]{\Christopher}
\affil[30]{\Hue}
\affil[31]{\Argonne}
\affil[32]{\PuertoRico}
\affil[33]{\VirginiaMilitary}
\affil[34]{\Syracuse}
\affil[35]{\Glasgow}
\begin{abstract}
The small binding energy of hypertrition leads to predictions of non-existence of bound hypernuclei for the isotriplet three-body system such as $nn\Lambda$.
However, invariant mass spectroscopy at GSI reported events that may be interpreted as the bound $nn\Lambda$ state.
The $nn\Lambda$ state was searched for by missing-mass spectroscopy via the $(e,e'K^+)$ reaction at the Jefferson Lab's experimental Hall A. 
The present experiment has higher sensitivity to the $nn\Lambda$-state investigation in terms of the better precision by a factor of about three than the previous invariant mass spectroscopy.
The analysis shown in this article focuses on the derivation of the reaction-cross section for the $^3{\rm{H}}(\gamma^{*},K^+)\textrm{X}$ reaction.
Events that were detected in an acceptance, where a Monte Carlo simulation could reproduce the data well ($|\delta p/p| < 4\%$), were analyzed to minimize the systematic uncertainty.
No significant structures were observed with the acceptance cuts, and the upper limits of the production-cross section of $nn\Lambda$ state were obtained to be $21$ and $31~\rm{nb/sr}$ at the $90\%$-confidence level when theoretical predictions of $(-B_{\Lambda}, \Gamma) = (0.25,0.8)$ and $(0.55, 4.7)$ MeV, respectively, were assumed.
The cross-section result provides valuable information to examine the existence of $nn\Lambda$.

\end{abstract}

\subjectindex{D14}
\maketitle

\section{\label{sec:Introduction} Introduction}
An extension of the nuclear (NN) interaction to the baryon (BB) interaction with strangeness degrees of freedom in the SU(3) flavor symmetry is one of the major subjects of study in nuclear physics.
Studies of hypernuclei which have at least one strange quark have played an important role in the understanding of the hyperon-nucleon (YN) interaction.

Hypertriton, which is composed of a neutron, a proton, and a $\Lambda$, is known as the simplest bound hypernulcear system.
The hypertrition binding energy was measured to be $130\pm50~\rm{keV}$ by an emulsion experiment~\cite{Juric1973}.
The small binding energy in the isospin singlet ($T=0$), three-body ($A=3$) hypernuclear system suggests that no bound systems exist for the isotriplet ($T=1$), $A=3$ hypernuclei such as $pp\Lambda$ and $nn\Lambda$.
However, events that may be interpreted as the bound state of $nn\Lambda$ were observed in a $t+\pi^-$ invariant mass spectroscopy at GSI~\cite{Rappold2013}.
The measured binding energy and width are  $-B_\Lambda=0.5\pm1.1\pm2.2~$~MeV and $\Gamma=5.4\pm1.4$~MeV, respectively for the vertex cut of $-2<Z<30$~cm.

As well as the early study by Downs and Dalitz~\cite{Downs1959}, theoretical calculations with the recently developed interaction models were unable to reproduce such a bound state that was observed at GSI.
For example, various methods such as Faddeev~\cite{Miyagawa1995,Afnan2015,Filikhin2016,Gal2014,Kamada2016,Garcilazo2014} and variational methods~\cite{Hiyama2014} do not support the occurrence of the bound state.
Although a calculation based on the pion-less effective field theory (\bcancel{$\pi$} EFT) suggested the existence of the bound state as an Effimov state, there were no quantitative discussions~\cite{Ando2015}.

On the other hand, the existence of a resonant state of $nn\Lambda$ was predicted by some theories.
By using two-body S-wave potential in the framework of the three-body Jost functions, Belyaev {\it{et al.}} suggested the existence of the resonant state, which according to their estimate had a decay width of approximately a few MeV~\cite{Belyaev2008}.
Filikhin {\it{et al.}} predicted a broader state~\cite{Filikhin2016} while Kamada {\it{et al.}} obtained a narrower state of less than $1~\rm{MeV}$~\cite{Kamada2016} by using the Faddeev-calculation method.
The \bcancel{$\pi$}-EFT calculation with specific interaction models also supported the existence of the broader state~\cite{Martin2020,Martin2021}.
A prediction by Afnan {\it{et al.}} suggested that even though a resonant state of $nn\Lambda$ does not exist with the standard strength of the $\Lambda n$ interaction, the existence of the resonant state may become possible with a more attractive $\Lambda n$ interaction by about $5~\%$~\cite{Afnan2015}.

The $nn\Lambda$ state needs to be experimentally confirmed.
In emulsion experiments, neutron rich systems such as $nn\Lambda$ are relatively hard to uniquely identify, and indeed, no signals of the $nn\Lambda$ system were found.
Conventional counting experiments that use the $(\pi^+,K^+)$ or the $(K^-,\pi^-)$ reactions never have an available target $(nnn)$ to produce such a state because these reactions convert a neutron into a $\Lambda$. 
Invariant mass spectroscopy with heavy ion beams or collisions can search for the signal of $nn\Lambda$, as was previously reported by HypHI Collaboration at GSI~\cite{Rappold2013}.
The invariant mass spectroscopy of the hypernuclei measures weak-decayed particles like the emulsion experiments.
Therefore, both the invariant mass spectroscopy and the emulsion experiment have much less sensitivity for the detection of the resonant state, which  tends to decay in the presence of the strong interaction.

Missing-mass spectroscopy is a powerful tool for the investigation of the $nn\Lambda$ state because it allows a measurement for both the bound and resonant states.
The missing-mass spectroscopy of $\Lambda$ hypernuclei with the $(e,e'K^+)$ reaction was developed at the Jefferson Lab (JLab), and the high resolution hypernuclear data were successfully published~\cite{Nue2013,Tang2014,Gogami2018,Franco2019, Gogami2021}.
However, a tritium target which requires strict regulations to be followed concerning safety issues, is necessary for the $(e,e'K^+)$-reaction spectroscopy, because a proton gets converted into a $\Lambda$ in the reaction.
The use of the tritium target was realized in 2017, thanks to the great efforts of JLab Tritium Target Group~\cite{Santiesteban2019}, which gave us a chance to perform an experiment for investigating the $nn\Lambda$ state at JLab's experimental Hall A (JLab E12-17-003 Experiment).

We performed the experiment with the $^3{\rm{H}}(e,e'K^+)\textrm{X}$ reaction in Oct--Nov, 2018.
Both arms of the high resolution spectrometers HRSs (HRS-L and HRS-R)~\cite{Alcorn2004} in JLab Hall A were used for the analyses of the momentum vectors of $e'$ and $K^+$ at reaction points to reconstruct the missing mass.
The data were successfully taken, and an analysis that focused on the cross-section derivation was performed.
Strict event-selection conditions particularly for the momentum selection were set for the present analysis to minimize the systematic uncertainty on the result. The event-selection conditions for the present analysis were set stricter than those of other undergoing analyses, such as
(i) a spectrum analysis with loosened cuts for peak search and (ii) a distribution analysis of the quasi-free $\Lambda$ (QF) production for a study of $\Lambda n$ final state interaction (FSI). 
These analyses (i and ii shown above) will be discussed elsewhere.

The present article contains the following details:
Our experimental kinematics for the $(e,e'K^+)$ reaction, the electron beam provided by Continuous Electron Beam Accelerator Facility (CEBAF) at JLab, and the experimental apparatus are discussed in Sec.~\ref{sec:Experiment}. 
Section~\ref{sec:Analysis} describes the data analysis.
Section~\ref{sec:Result} discusses the result and discussion followed by a conclusion in Sec.~\ref{sec:Conclusion}.

\section{\label{sec:Experiment} Experiment}
The missing-mass spectroscopy measurements using the $(e,e'K^+)$ reaction were conducted at JLab Hall A.
In this experiment, $4.32$-${\rm{GeV}}/c$ electron beams were impinged on the tritium target. The scattered electrons and the produced $K^+$'s were measured using the two spectrometers, HRS-L and HRS-R, respectively~\cite{Alcorn2004}.

\subsection{\label{sec:Kinematics} Kinematics}
\begin{figure}[bt]
\centering
\centering
\includegraphics[scale=0.35]{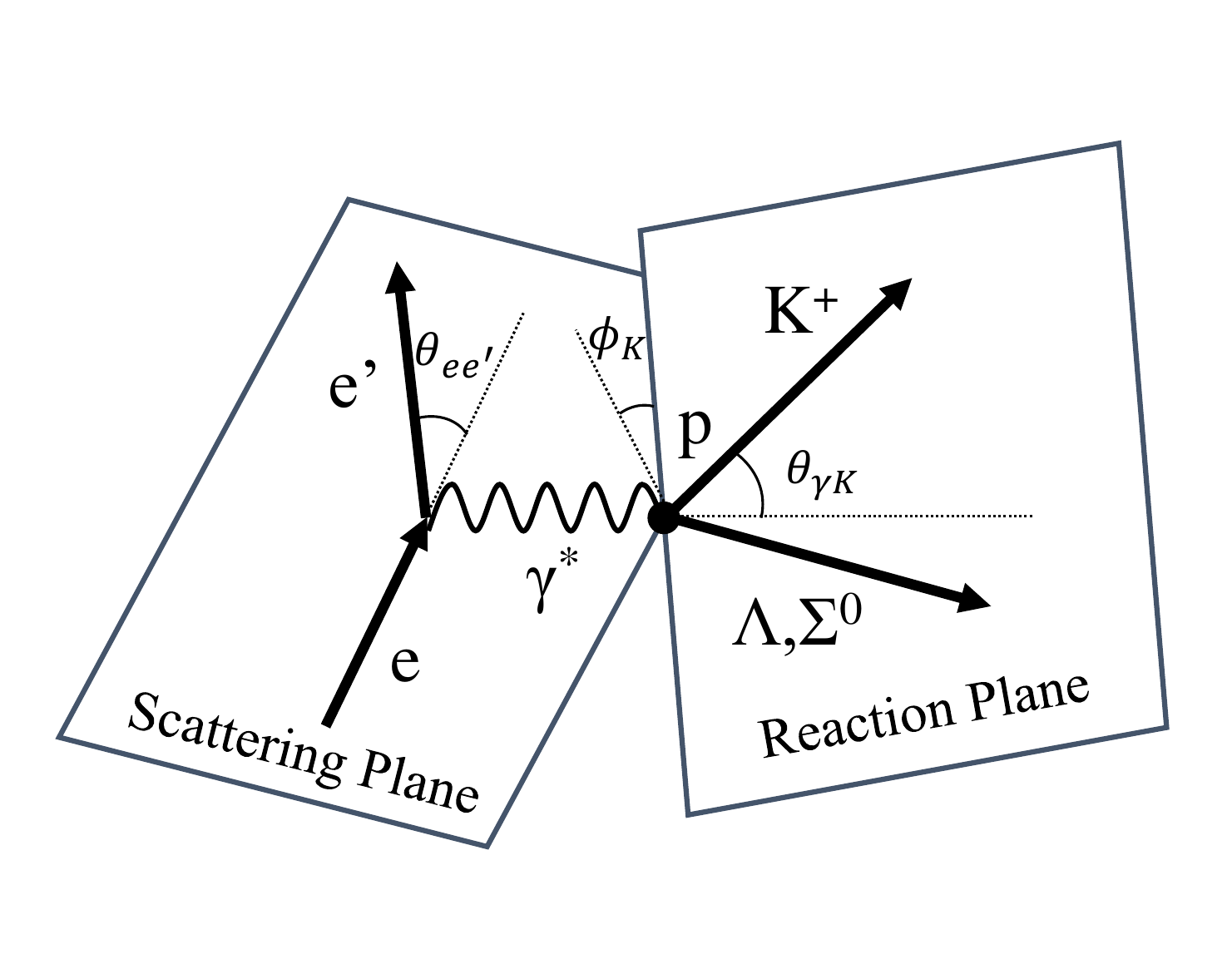}
\caption{\label{fig:kinematics} Schematics of the $(e,e'K^+)$ reaction. A virtual photon reacts with a proton to produce the $\Lambda$ (or $\Sigma^0$) and a $K^+$. }
\end{figure}

A schematic of the $(e,e'K^+)$ reaction is shown in Fig.~\ref{fig:kinematics}.
The one-photon exchange approximation, which assumes that a virtual photon mediates the reaction, is generally used in the electro-production.

The energy and momentum of the virtual photon are defined as follows,
\begin{eqnarray}
\omega &=& E_e - E_{e'}, \\
\vec{q} &=& \vec{p}_e - \vec{p}_{e'},
\end{eqnarray}
where the four momenta of the incident electron and the scattered electron are denoted by $(E_{e,e'}, \vec{p}_{e,e'})$.
Similarly, $(\omega, \vec{q})$ stands for the four momentum of the virtual photon.
The triple differential cross section for the hyperon production may be described as

\begin{eqnarray}
{ 
{\frac{d^3\sigma}{{dE_{e'}}{d\Omega_{e'}}{d\Omega_K}} }}
= \Gamma \left[\frac{d\sigma_U}{d\Omega_K}
+ \epsilon_L\frac{d\sigma_L}{d\Omega_K}
+\epsilon \frac{d\sigma_P}{d\Omega_K} \right.
\left.+\sqrt{\epsilon_L(1+\epsilon)}\frac{d\sigma_I}{d\Omega_K} \right],
\end{eqnarray}
where $\sigma_U$, $\sigma_L$, $\sigma_P$ and $\sigma_I$ are the terms of the unpolarized transverse, longitudinal, polarized transverse and interference cross sections, respectively~\cite{Adam1992,Sotona1994,Hungerford1994}.
$\Gamma$ represents the virtual photon flux, which is defined as
\begin{eqnarray} \label{VirtualPhotonFlux}
\Gamma = \frac{\alpha}{2\pi^2Q^2} \frac{E_\gamma}{1-\epsilon}
\frac{E_{e'}}{E_e},
\end{eqnarray}
where $\alpha$ is the fine structure constant, $E_\gamma~(= \omega + q^2/2m_p)$ is the effective photon energy, and $Q^2~(=-q^2)$ is the square of the four momentum transfer, with the negative sign.
The transverse polarization $\epsilon$ and the longitudinal polarization $\epsilon_L$ are defined as follows,
\begin{eqnarray}
\epsilon &=&\left( 1+\frac{2|\vec{q}|^2}{Q^2}\tan^2\frac{\theta_{e'}}{2}  \right)^{-1}, \\
\epsilon_L &=& \frac{Q^2}{\omega^2}\epsilon.
\end{eqnarray}

\begin{figure}[bt]
\centering
\includegraphics[scale=0.4]{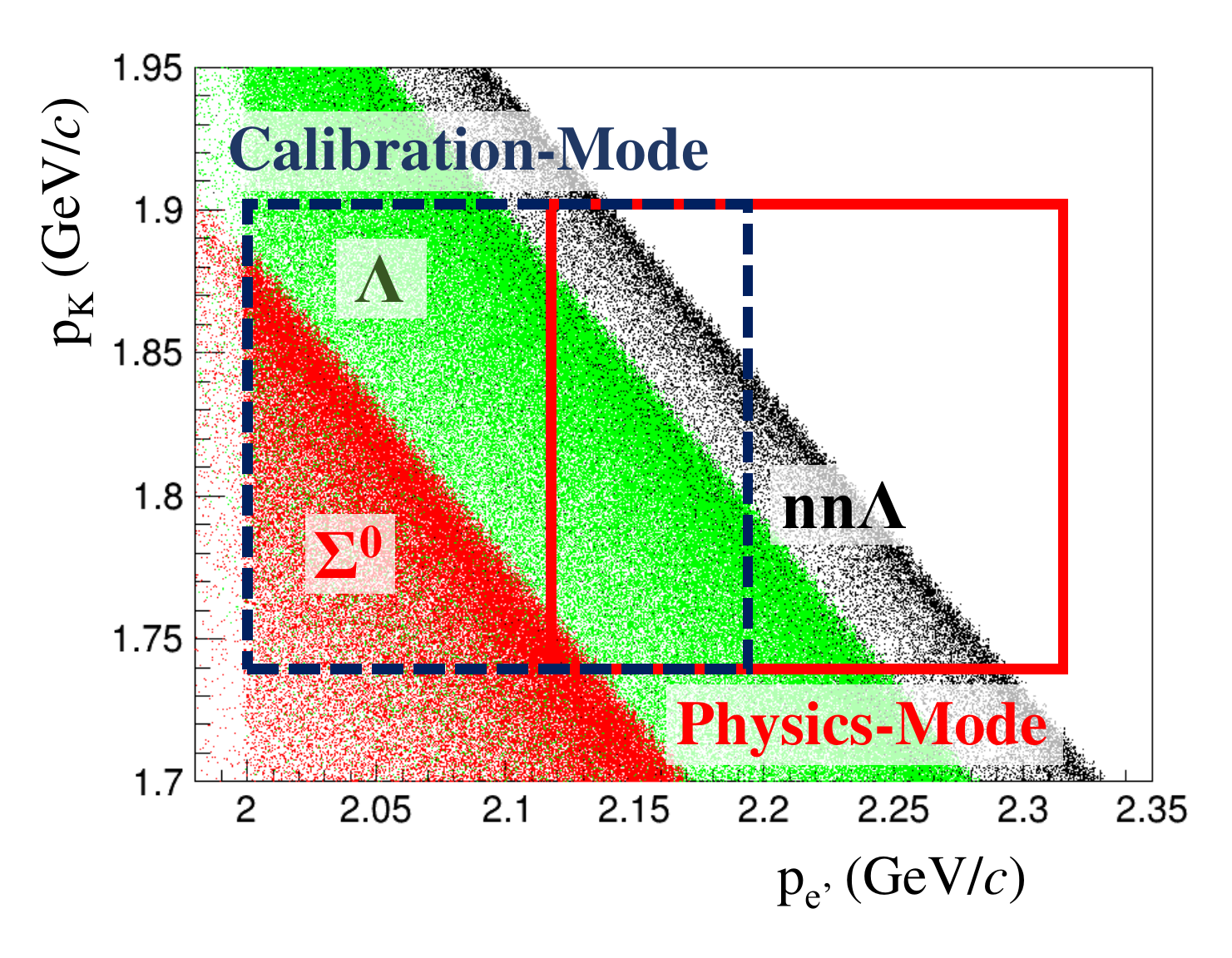}
\caption{\label{fig:kine2D} (Color Online) Correlations between the momenta of $e'$ and $K^+$ for the $p(e,e'K^+)\Lambda$ (green dots), $p(e,e'K^+)\Sigma^0$ (red dots), and ${\rm{^3H}}(e,e'K^+)nn\Lambda$ (black dots) reactions in a Monte Carlo simulation. Boxes with solid and dashed lines represent acceptances for the physics and the calibration modes (M$_{\rm{{phys.}}}$ and M$_{\rm{{calib.}}}$), respectively.}
\end{figure}

The HRSs were at the scattering angles of $\theta_{ee'} = \theta_{eK} = 13.2^{\circ}$ in the laboratoly frame and had central momenta of $2.218$~GeV/{\it{c}} and $1.823$~GeV/{\it{c}} for $e'$ and $K$, respectively for the physics runs (physics mode: $\rm{M_{phys.}}$).
The mode $\rm{M_{phys.}}$ had a sufficient acceptance to cover the region where the $nn\Lambda$ state may exist.
A proton (hydrogen) target was also used in the mode $\rm{M_{phys.}}$ to measure the $p(e,e'K^+)\Lambda$ reaction that was used for the energy calibration, as shown in Sec.~\ref{sec:Missing-mass reconstruction}.
For the purpose of calibration, data with a different momentum setting for the $e'$ (calibration mode: M$_{\rm{calib.}}$) was used. 
In the mode M$_{\rm{calib.}}$, the $e'$ central momentum was decreased to $2.100~\rm{GeV/\it{c}}$, which allowed simultaneous measurements of $\Lambda$ and $\Sigma^0$ from the proton target, whereas the $nn\Lambda$ production became almost out of the acceptance (Fig.~\ref{fig:kine2D}).
It may be noted that we did not decrease the $K^+$ momentum to avoid the low survival probability of the $K^+$.
The kinematical parameters are summarized in Tab.~\ref{tab:kinematics}.

\begin{table}
\centering
\caption{Major parameters for the present experiment. The $p(e,e'K^+)\Lambda$ reaction was assumed for the calculations of $\sqrt{s}$ and the momentum transfer $q$.}
\label{tab:kinematics}
\begin{tabular}{lcc}
\hline
\textrm{}&
\textrm{Calibration Mode}&
\textrm{Physics Mode } \\
 &  ($\rm{M_{calib.}}$) & ($\rm{M_{phys.}}$) \\
\hline
Reaction & $p(e,e'K^+)\Lambda/\Sigma^0$ & 
\begin{tabular}{c}
$p(e,e'K^+)\Lambda$ \\ $^3{\rm{H}}(e,e'K^+)nn\Lambda$
\end{tabular} \\ 
\hline
$p_{e'}^{\rm{cent.}}~(\rm{GeV}/\it{c})$ & 2.100 & 2.218 \\
$p_K^{\rm{cent.}}~(\rm{GeV/\it{c}})$ & \multicolumn{2}{c}{1.823} \\
$Q^2~(\rm{GeV/\it{c}})^2$ & 0.479 & 0.505\\
$\theta_{e\gamma}$~(deg) & 11.9 & 13.2 \\
$q~(\rm{GeV/\it{c}})$ & 0.497 & 0.389 \\
$\sqrt{s}~(\rm{GeV})$ & 2.13 & 2.07 \\
$\epsilon$ & 0.769 & 0.794 \\
$\epsilon_L$ & 0.075 & 0.092 \\
\hline
\end{tabular}
\end{table}

\subsection{\label{sec:Electron beam} Electron beam}
We used $4.32$-$\rm{GeV/\it{c}}$ electron beams that were provided by CEBAF at JLab.
The typical beam current on the target was $22.5~\mu$A with a raster size of $2\times2~\rm{mm^2}$. 
The spread and the drift of the beam energy were well controlled and monitored during the experiment~\cite{Santiesteban2021}.
The total energy uncertainty was approximately $\Delta E/E \lesssim 1\times10^{-4}$ in FWHM.
The beam current was measured by the Parametric Current Transformer system and the beam current monitor, with an uncertainty of approximately $1.0\%$~\cite{Unser1969,Unser1981,Santiesteban2019}.
The total beam charge on the tritium and hydrogen target were $16.9$ and $4.7~\rm{C}$, respectively, for the data used in the present analysis.

\subsection{\label{sec:Overview of experimental apparatus} Overview of experimental apparatus}
The tritium gas was enclosed in a target cell made with an Al alloy Al-7075.
The target-cell length along the beam direction was $25~\rm{cm}$, and the areal density was $84.8\pm0.8~\rm{mg/cm^2}$ for the gas.
The target cell was cooled down to $40~\rm{K}$ during the beam operation resulting in a gas pressure of $0.3$~MPa.
The hydrogen gas was used for the energy calibration as shown in Sec.~\ref{sec:Missing-mass reconstruction}.
The target cell for the hydrogen gas was of the same shape as that for the tritium target.
The areal density for the hydrogen target was $70.8\pm0.4~\rm{mg/cm^2}$.
The gas density was reduced during the beam irradiation due to a local heat deposit along the beam path.
The gas-density reduction was evaluated as a function of the beam intensity, and it was found that the tritium-gas density is reduced by $10\%$ at the beam intensity of $22.5~\mu$A~\cite{Santiesteban2019}.
It is noted that the tritium decays to $^3\rm{{He}}$ with a lifetime of $12.32$~years~\cite{Firestone2016}.
The reduction effect of the tritium nuclei due to the decay was taken into account when the cross section was calculated, as shown in Sec.~\ref{sec:Derivation of the differential cross section}.

HRS-L and HRS-R were used for the detection of the $e'$ and $K^+$, respectively.
Each spectrometer is composed of three quadrupole and one dipole magnets~(QQDQ).
The optical features of the HRSs are basically identical.
The path length from the target to the focal plane is $23.4~\rm{m}$. 
The designed momentum resolution is $\Delta p/p = 1\times10^{-4}$~(FWHM).
However, the momentum resolution was limited because of the materials, particularly for the target cell.
An expected energy resolution in a resulting spectrum for which the effects from the target cell material etc. were taken into account is described in Sec.~\ref{sec::Energy Resolution}.

The configurations of the spectrometers were similar to each other.
Each spectrometer had vertical drift chambers (VDC) for the particle tracking~\cite{Fissum2001} and plastic scintillation detectors (S0 and S2) for the time-of-flight measurement, which were installed in this order from the upstream.
Cherenkov detectors were installed between S0 and S2 for the particle identification~\cite{Marrone2008}.
The plastic scintillation detectors were used for a data-taking trigger with the following condition: 
$\rm{(S0\otimes S2)_L \otimes (S0\otimes S2)_R}$, 
where the subscripts L and R represent the hit conditions for HRS-L and HRS-R, respectively.
The Cherenkov detectors were not used for the main trigger but were used in the off-line analyses.
A $\rm{CO_2}$-gas Cherenkov detector was used to remove $\pi^-$'s in the HRS-L.
In HRS-R, on the other hand, two of the aerogel-Cherenkov detectors (AC1 and AC2; refractive indices of $n=1.015$ and $1.055$, respectively) were used to remove the background $\pi^+$'s and protons as shown in Sec.~\ref{sec:Particle identification}.

\subsection{\label{sec:Summary of measurements} Summary of measurements}
We performed missing-mass spectroscopy with the ${^3\rm{H}}(e,e'K^+)\textrm{X}$ reaction at the JLab Hall A.
Two existing spectrometers (HRSs) were used to detect the $e'$ and $K^+$. 
The ${\rm{H}}(e,e'K^+)\Lambda$ and ${\rm{H}}(e,e'K^+)\Sigma^0$ reactions were also measured for the purpose of calibration.
The experimental data were taken in Oct--Nov, 2018.

\section{\label{sec:Analysis} Analysis}

\subsection{\label{sec:Missing-mass reconstruction} Missing-mass reconstruction}
The missing mass was reconstructed using the following equation,
\begin{eqnarray} \label{eq:Missing}
{\rm M}_{X}=
\sqrt{(E_e+{\rm M}_{\rm{t}}-E_K-{E_{e'}})^2 
- (\vec{p_e} - \vec{p_K} - {\vec{p_{e'}}})^2 }. 
\end{eqnarray}
The beam energy was precisely determined by CEBAF.
Therefore, to reconstruct the missing mass, the momentum ($p_{\rm{T}}$), and angles ($x'_{\rm{T}}=p_x/p_z$ and $y'_{\rm{T}}=p_y/p_z$) of $e'$ and $K^+$ at the production point were necessary.
The momentum and angles were obtained from the following polynomial functions,
\begin{eqnarray}
x'_{\rm{T}} &=& \sum_{a+b+c+d+e=0}^n C_{x'}(a,b,c,d,e)x^ax'^by^cy'^dz_{\rm{T}}^e \label{TMtheta},\\ 
y'_{\rm{T}} &=& \sum_{a+b+c+d+e=0}^n C_{y'}(a,b,c,d,e)x^ax'^by^cy'^dz_{\rm{T}}^e \label{TMphi}, \\
p_{\rm{T}} &=& \sum_{a+b+c+d+e=0}^n C_{p}(a,b,c,d,e)x^ax'^by^cy'^dz_{\rm{T}}^e \label{TMp},
\end{eqnarray}
where $x$, $y$, $x'$, and $y'$ are positions and angles at the focal plane.
$z_{\rm{T}}$ is a production position along the beam direction, which was obtained as follows,
\begin{eqnarray}
z_{\rm{T}} &=& \sum_{a+b+c+d=0}^n C_{z}(a,b,c,d)x^ax'^by^cy'^d. \label{TMz}
\end{eqnarray}
$C_{x',\ y',\ p,\ z}$ in Eqs.~(\ref{TMtheta})--(\ref{TMz}) are the parameters to be optimized by using various calibration data as described below.
We took $n=5$ and $4$ for the momentum and angle functions, respectively.
In Eqs.~(\ref{TMtheta})--(\ref{TMp}), the power of $z_{\rm{T}}$ was limited to $e\leq 2$.
In Eq.~(\ref{TMz}), $n=3$ was used.
These powers were set as small as possible to avoid over tuning, maintaining expected resolutions.

\begin{figure}[bt]
\centering
\includegraphics[scale=0.4]{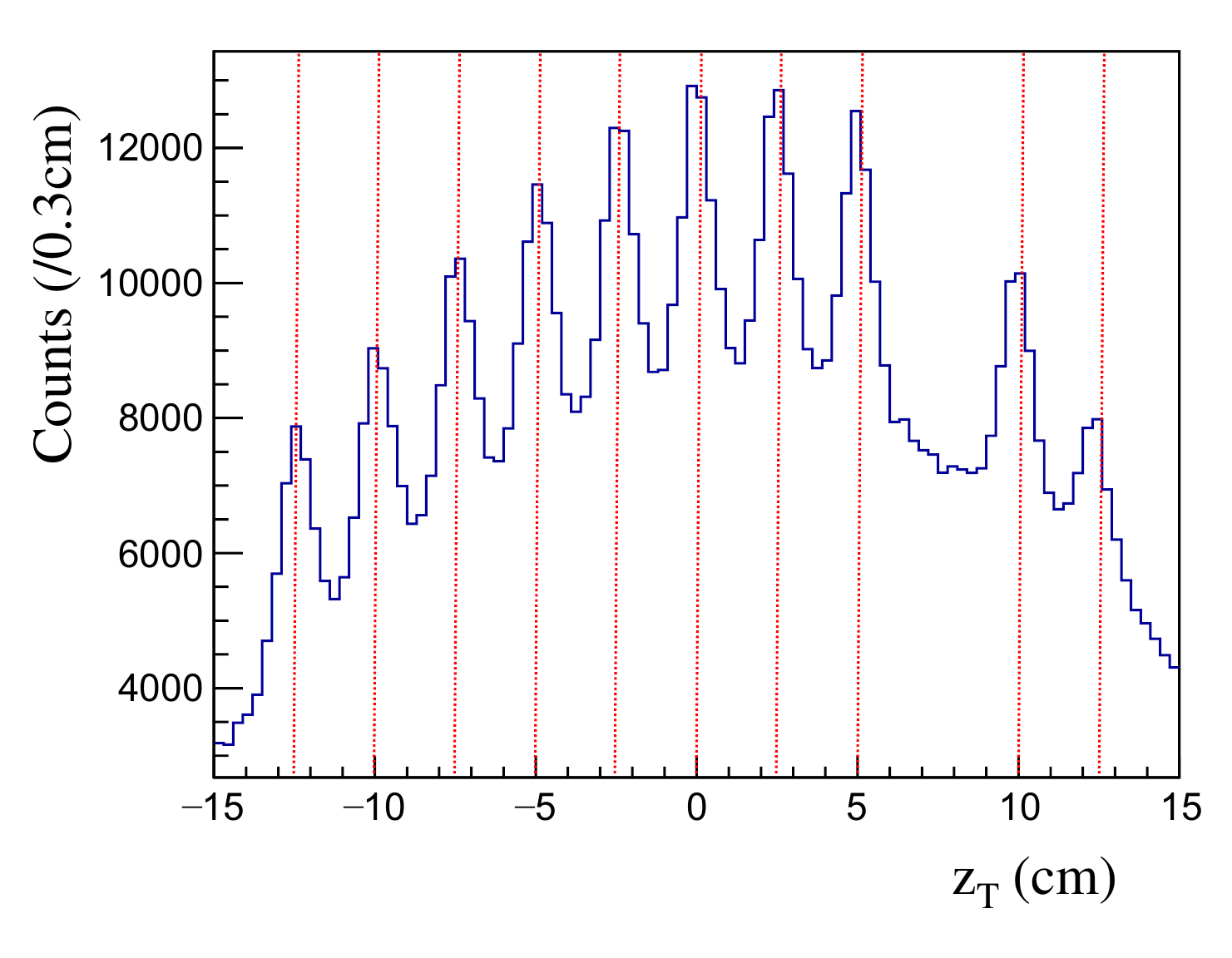}
\caption{\label{fig:foil} Reconstructed $z_{\rm{T}}$ for the data with the multi-carbon foils by using HRS-L.
} 
\end{figure}

Multi-carbon foils were used as targets instead of the gas target for the $z_{\rm{T}}$ calibration.
There were ten foils that were placed at a distance of $2.5~{\rm{cm}}$ from each other except for the second and third foils, between which the distance was set to $5~{\rm{cm}}$.
Each foil had an areal density of approximately $45~{\rm{mg/cm^2}}$.
Figure \ref{fig:foil} shows a reconstructed $z_{\rm{T}}$ distribution by HRS-L.
Separated peaks from the carbon foils are clearly seen.
The $C_{z}$ parameters in Eq.~(\ref{TMz}) for both of HRSs were optimized to reproduce the foil positions by using the MINUIT algorithm~\cite{James1975,Rene1997}.

\begin{figure}[bt]
\centering
\includegraphics[scale=0.4]{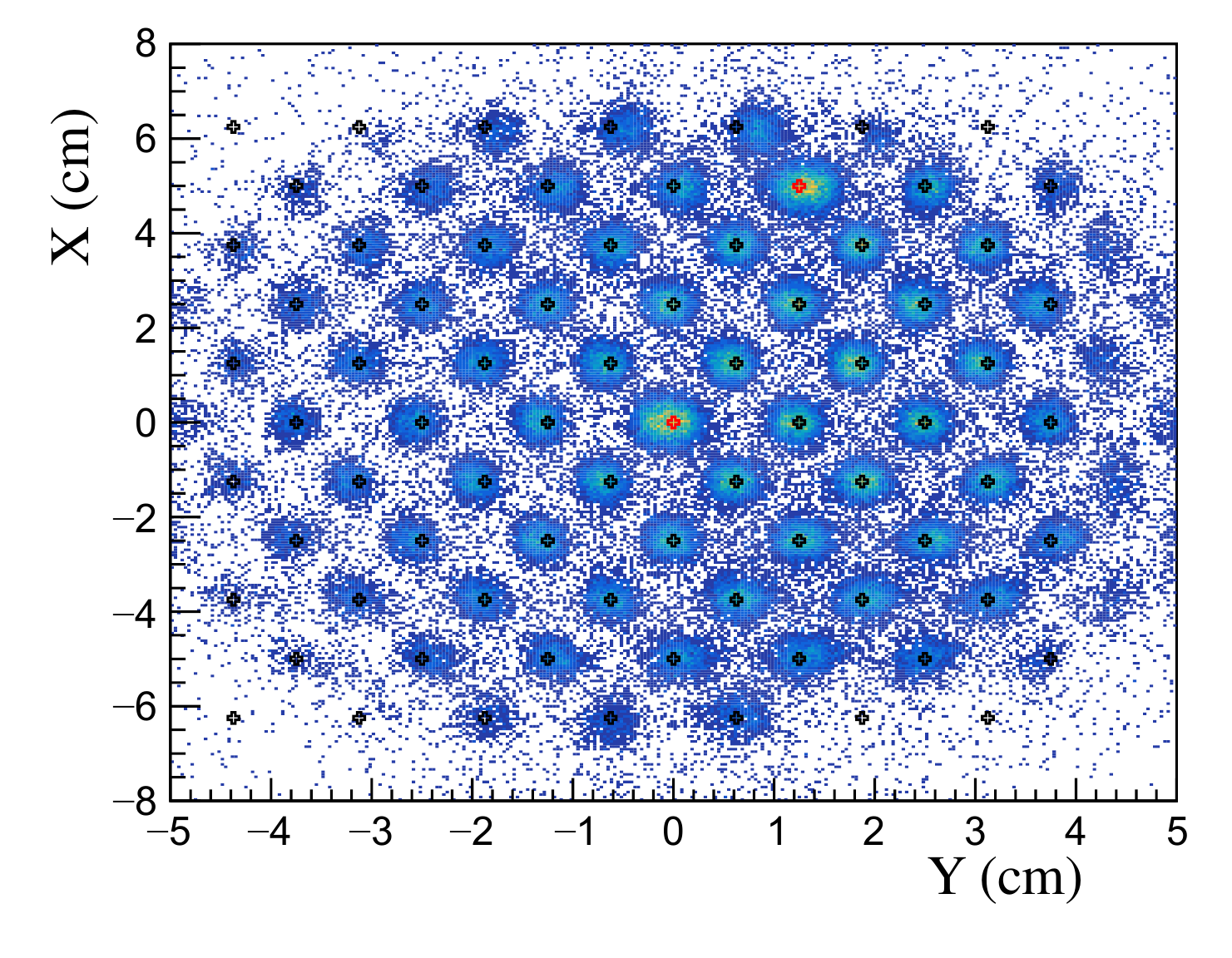}
\caption{\label{fig:sieve} $xy$ image at the sieve slit (SS) for the sieve slit data in HRS-L.
The image was reconstructed by using the reconstructed angles and the distance between the target and SS. Markers represent the hole positions that we expected.}
\end{figure}

The parameters of angles, which are $C_{x'}, C_{y'}$ in Eqs.~(\ref{TMtheta}) and (\ref{TMphi}), were optimized by using the calibration data with sieve slits.
The sieve slits (SS) were made of a $2.54$-cm thick tungsten plate provided with some holes, through which the particles can be detected in the spectrometers~\cite{Urciuoli2019}.
Dedicated data were taken with the SS attached in front of the first quadrupole magnet for each HRS.
Figure~\ref{fig:sieve} shows a particle-position image in HRS-L that was reconstructed by using the particle angles and the distance between the target and SS.
The angle parameters in Eqs.~(\ref{TMtheta}) and (\ref{TMphi}) were optimized to reproduce the hole patterns that were expected.
It may be noted that the holes had diameters of $4$ and $6$ mm.

The momentum parameters, which are $C_p$ in Eq.~(\ref{TMp}), were optimized by using the data obtained from the hydrogen target.
Events of $\Lambda$ and $\Sigma^0$ productions were detected in $\rm{M_{calib.}}$, whereas only $\Lambda$-production events were detected in $\rm{M_{phys.}}$.
The missing mass was reconstructed to observe the $\Lambda$ and the $\Sigma^0$ peaks. The $C_p$ were optimized to make their peak means consistent with the PDG masses~\cite{PDG2020}.
Missing-mass resolution after the momentum calibration was found to be $\sigma = 1.3\pm0.1~\rm{MeV/}{\it{c}}^2$ when a fit was performed with a Gaussian function over a range of $|{\rm M}_X - {\rm M}_{\Lambda}|<2$~MeV/${\it{c}}^2$.

The $\Lambda$ binding energy is defined as follows: $B_\Lambda = {\rm M_{core}} + {\rm M_\Lambda} - {\rm{M}}_X$.
In the present analysis, we took ${\rm M_{core}} = 2{\rm M}_n$, where ${\rm{M}}_n$ is the mass of a neutron.
The masses of the $\Lambda$ and the neutron ${\rm{M}}_{\Lambda,n}$ were taken from Ref.~\cite{PDG2020}.
The nuclear masses of the hydrogen and tritium targets $\rm{M_t}$ used in Eq.~(\ref{eq:Missing}) were taken from the Refs.~\cite{PDG2020} and \cite{Wang2021}, respectively.

\subsection{\label{sec:Target}Event selection for gas target}
\begin{figure}[bt]
\centering
\includegraphics[scale = 0.4 ]{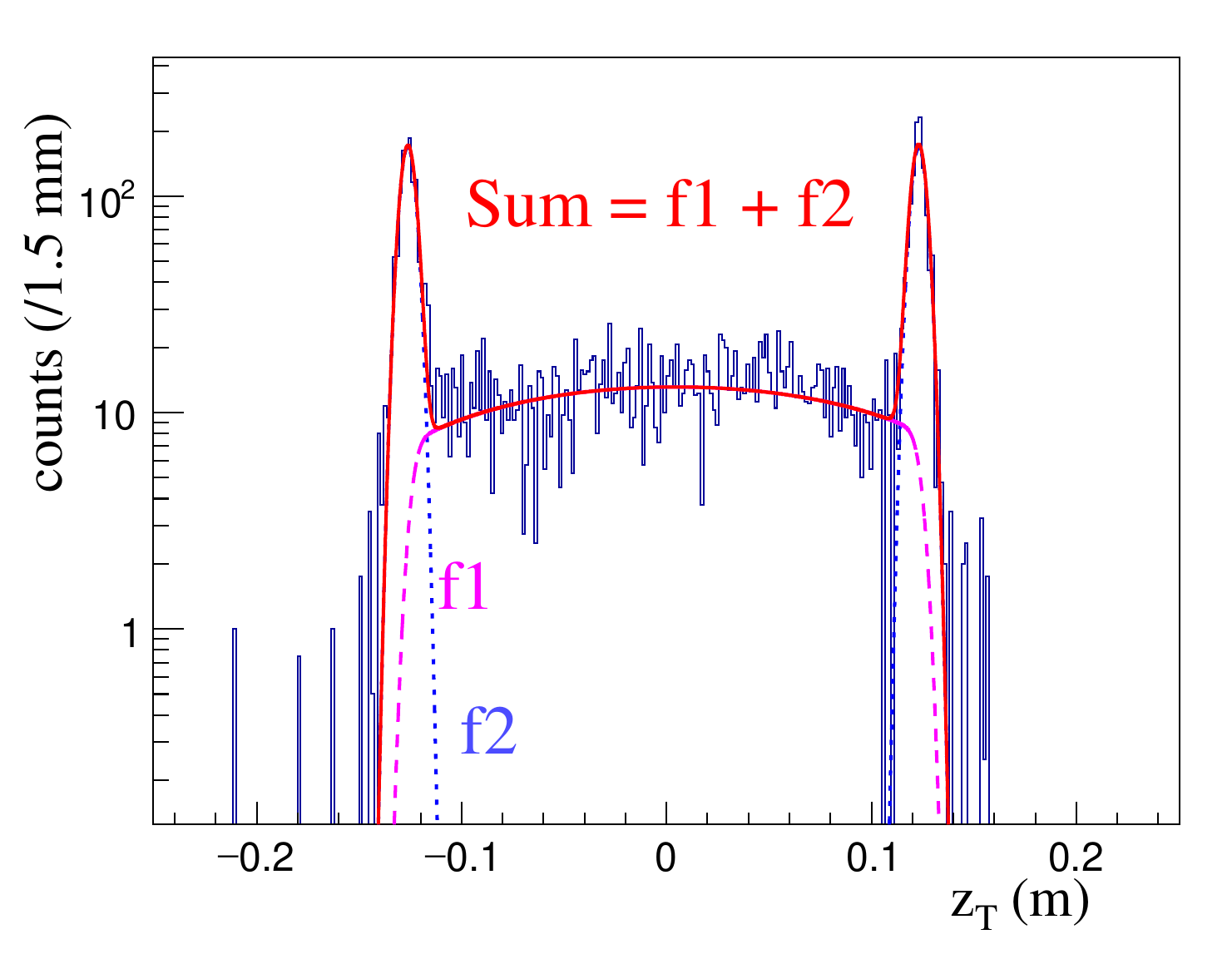}
\caption{\label{fig:target_survival} Distribution of $z_{\rm{T}}^{\rm{mean}} = (z^L_{\rm{T}} + z^R_{\rm{T}})/2$ for the data with the tritium target.
A fit with the 2nd-order polynomial function convoluted by the Gaussian function ($f_1$) and the two Gaussian functions ($f_2$) was performed for the gas and the target cell regions, respectively.
It is noted that the width of the Gaussian function of $f_1$ is the same as that of $f_2$.
}
\end{figure}

The production position along the beam axis $z_{\rm{T}}$ was independently reconstructed in HRS-L and HRS-R.
Event selection by the $z_{\rm{T}}$ difference between the HRS-L and HRS-R was applied as follows, $|z_{\rm{T}}^{\rm{diff}}| = |z_{\rm{T}}^{\rm{L}}-z_{\rm{T}}^{\rm{R}}|<2$~cm.
Figure~\ref{fig:target_survival} shows an average of $z_{\rm{T}}^{\rm{L}}$ and $z_{\rm{T}}^{\rm{R}}$; $z_{\rm{T}}^{\rm{mean}} = (z_{\rm{T}}^{\rm{L}}+z_{\rm{T}}^{\rm{R}})/2$.
The individual resolution of $\sigma(z_{\rm{T}}^{\rm{L}})=0.53$~cm and $\sigma(z_{\rm{T}}^{\rm{R}})=0.50$~cm were improved to $\sigma(z_{\rm{T}}^{\rm{mean}})=0.38$~cm by considering the average.
Peaks at $-12.5$ and $+12.5$~cm correspond to events from the target cell.
Therefore, the events of $|z_{\rm{T}}^{\rm{mean}}| < 10$~cm were selected for the gas-target analysis.
A fit was performed to estimate the amount of gas used for the analysis with the cut as shown in Fig.~\ref{fig:target_survival}.
The 2nd-order polynomial function convoluted by a Gaussian function $f_1$, and two Gaussian functions $f_2$ were used for the gas and the cell regions, respectively.
Here, the widths of Gaussian functions of $f_1$ and $f_2$ are the same.
As a result, $71\%$ of the full amount of gas was used with the cuts of $z^{\rm{diff}}$ and $z^{\rm{mean}}$.
The contamination from the target cell was estimated to be less than $0.1\%$, which is fairly small.

\subsection{\label{sec:Particle identification}Particle identification}
\begin{figure}[bt]
\centering
\includegraphics[scale =0.4]{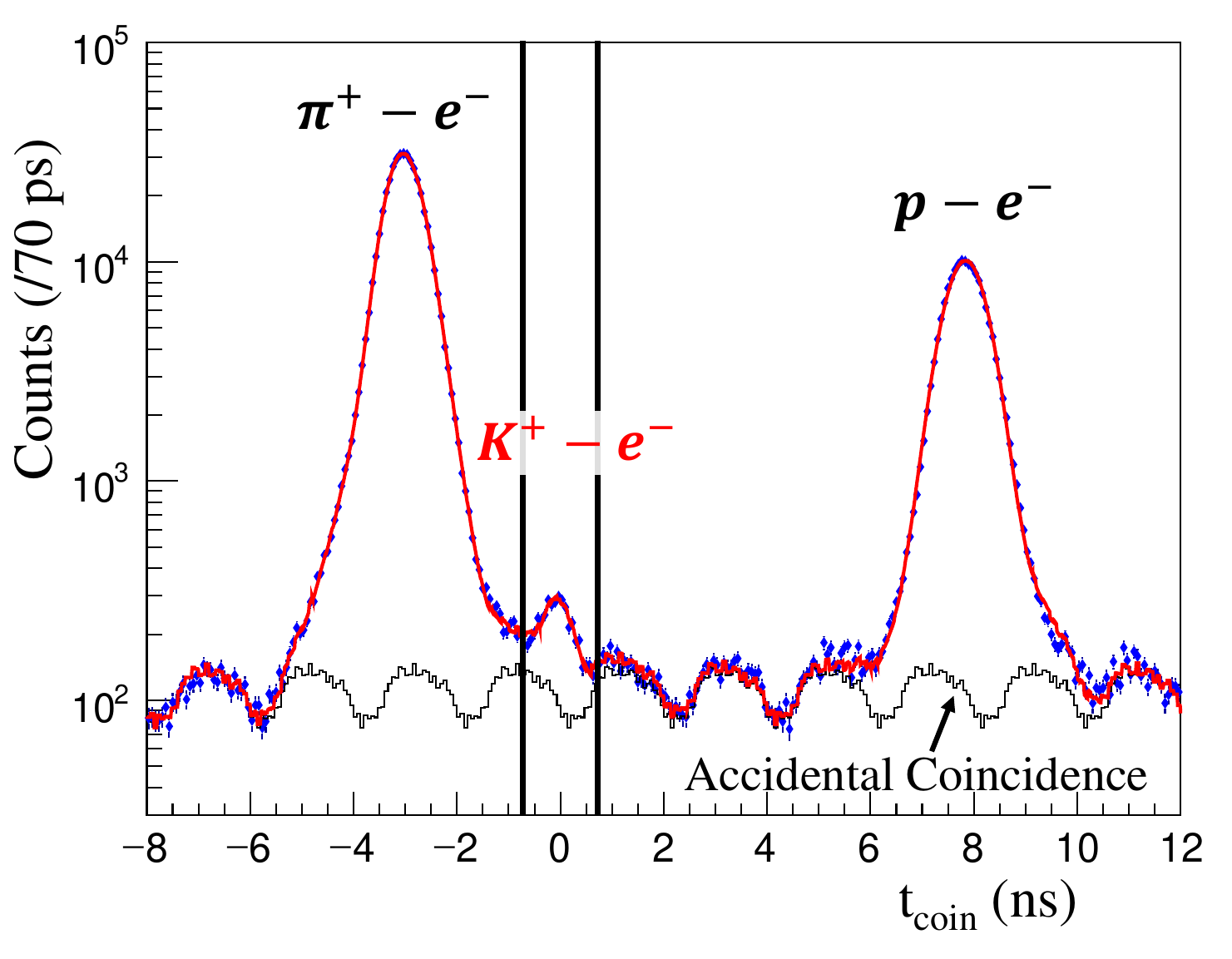}
\caption{\label{fig:coin} Distribution of coincidence time which is defined in Eq.~(\ref{eqs:tcoin}).
A distribution of the accidental coincidence backgrounds was evaluated by collecting some of accidental bunches. The fitting result is shown by the red solid line.
}
\end{figure}

The real coincidence event between the $e'$ and $K^+$ was selected by a selection of coincidence time.
The coincidence time was defined as follows,
\begin{eqnarray} \label{eqs:tcoin}
t_{\rm{coin}} = t_{\rm{t}}^{\rm{R}} - t_{\rm{t}}^{\rm{L}},
\end{eqnarray}
where $t_{\rm{t}}^{\rm{L,R}}$ are times at the target.
The times $t_{\rm{t}}^{\rm{L,R}}$ were obtained by using the reconstructed momenta, path lengths from the target to the scintillation detectors, and times at the scintillation detectors, and by assuming the masses of $e'$ and $K^+$, respectively.
Figure~\ref{fig:coin} shows the $t_{\rm{coin}}$.
A peak for the real coincidence of $e'$-$K^+$ is found at zero.
However, the other coincidence events such as $e'$-$\pi^+$ and $e'$-$p$ were found at the different times, which happened due to the wrong assumptions of the masses of $e'$ and $K^+$.
The peak width for the $K^+$ is approximately $300$~ps, and events of the $|t_{\rm{coin}}|<0.7$~ns were selected for the analysis.
The accidental coincidences are seen every $2$~ns, which is consistent with a beam-bunch cycle.
The accidental background distribution under the real coincidence events was evaluated from the data by collecting some of the accidental peaks and is shown in Fig.~\ref{fig:coin}.

\begin{figure}[bt]
\centering
\includegraphics[scale=0.4]{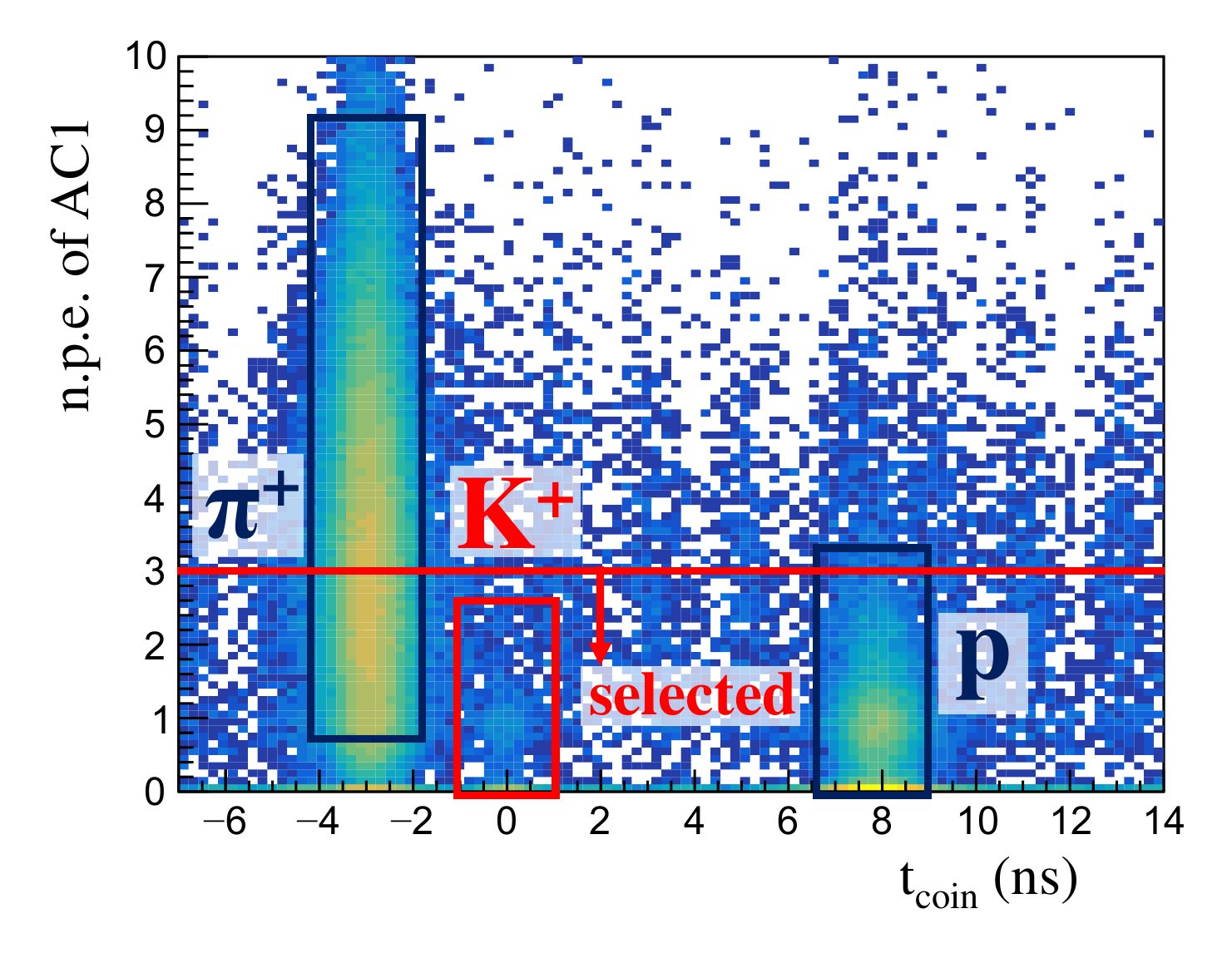}
\caption{\label{fig:AC1} The number of photoelectrons in AC1 as a function of $t_{\rm coin}$.}
\end{figure}

\begin{figure}[bt]
\centering
\includegraphics[scale=0.4]{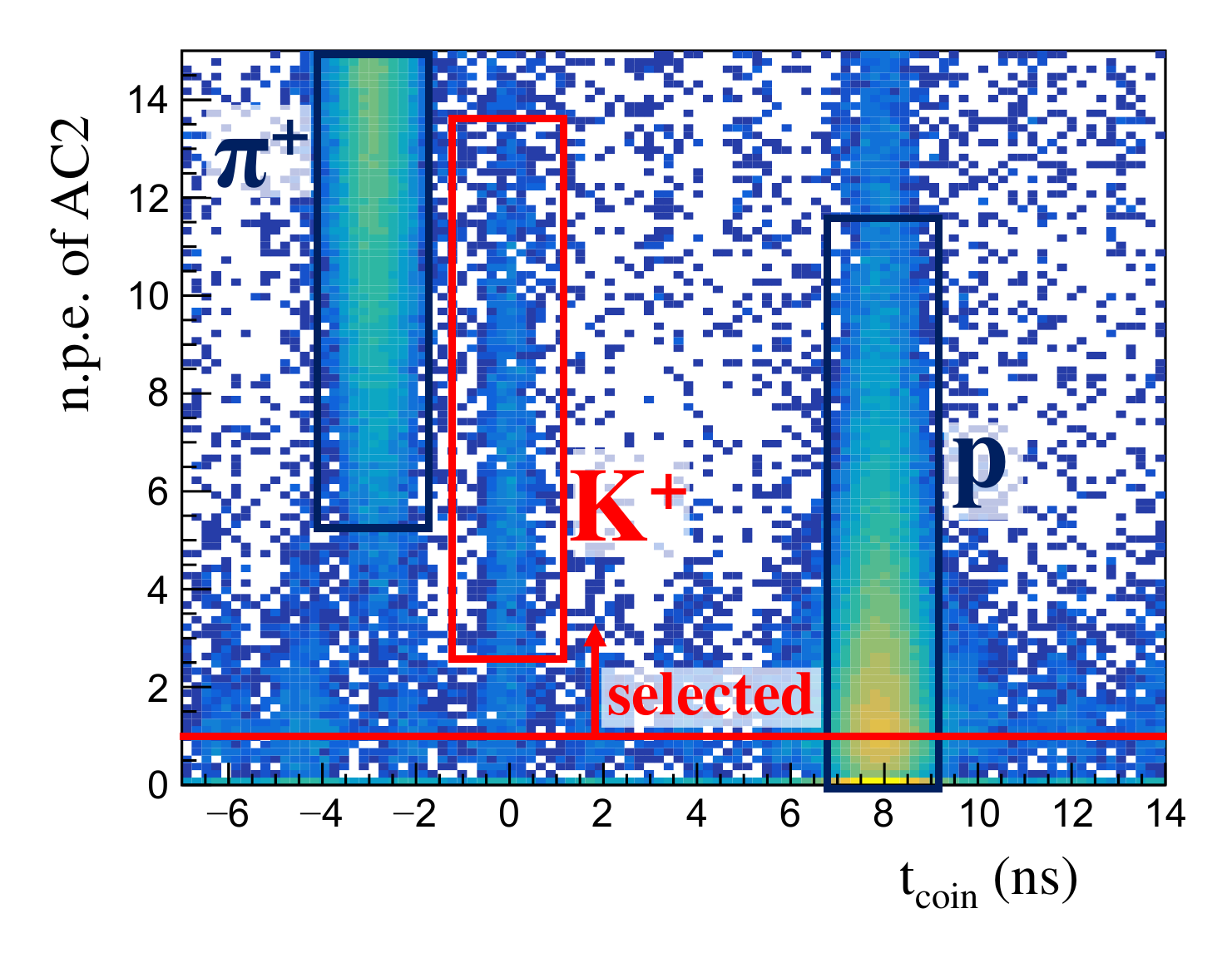}
\caption{\label{fig:AC2} The number of photoelectrons in AC2 as a function of $t_{\rm coin}$.}
\end{figure}

Major backgrounds in HRS-L are $\pi^-$'s, and they were rejected by the $\rm{CO_2}$-gas Cherenkov detector.
On the other hand, major backgrounds of protons and $\pi^+$'s in HRS-R were suppressed by light yield selections of the aerogel-Cherenkov detectors, AC1 and AC2.
The AC1 yields the Cherenkov light for $\pi^+$ whereas the AC2 yields the Cherenkov light for $\pi^+$ and $K^+$ in the momentum acceptance.
Figure~\ref{fig:AC1} and \ref{fig:AC2} show the number of photoelectrons (n.p.e.) in AC1 and AC2 as a function of $t_{\rm{coin}}$, respectively.
There are clear differences in the n.p.e. depending on the particle types.
We selected the $K^+$ by applying the following cuts,
\begin{eqnarray}
{\rm n.p.e.}~({\rm{AC1}}) &<& 3.0, \nonumber \\
1 < {\rm n.p.e.}~({\rm{AC2}}) &<& 23.0. \nonumber
\end{eqnarray}
$91.4 \pm 6.3\%$ of $e'$-$K^+$ coincidence events survived with these cuts in addition to the gas-Cherenkov cut.
The fraction of $\pi^+$ contamination was evaluated to be $2.4\pm1.8\%$ relative to the events that were identified as $K^+$.
The proton fraction was negligibly small.

\subsection{\label{sec:Monte Carlo simulation} Monte Carlo simulation}
A Monte Carlo simulation based on Geant4~\cite{Agostinelli2003,Allison2006,Allison2016} was coded and used to estimate the acceptance and other factors required for the cross-section analysis such as the decay and absorption factors of $K^+$'s in HRS-R~(Sec.~\ref{Efficiencies}).
In addition, the MC simulation was used to estimate the momentum loss in materials such as the target, detectors, air, and so on~(Sec.~\ref{sec:Energy loss correction}).
In the MC simulation, the precise geometries were modelled.

A three dimensional magnetic-field map for the dipole magnet was calculated using Opera3D (TOSCA), and was incorporated into the MC simulator.
In contrast, magnetic fields for the quadrupole magnets were calculated by an empirical formula taken from Ref.~\cite{Kato2009}.
The magnetic fields obtained by the calculation were not the same as those for the real experiment because of an arithmetic precision and an imperfection of the model.
Therefore, we scanned the magnetic-field strengths to find reasonable magnetic field settings that reproduced the experimental data.
We scanned various combinations of magnets' field strengths (QQDQ), comparing distributions of momentum vectors at the production point and angle distributions of the particles at the focal plane between the simulation and the real data.
The effective strengths of the magnetic fields obtained by the scan were used for the MC simulation to estimate the acceptance, and some correction factors as shown in Sec.~\ref{Efficiencies}.
The systematic error on the final result originating from the magnetic field settings (acceptances) are described in Sec.~\ref{sec::Systematic uncertainties}.

\begin{figure}[bt]
\centering
\includegraphics[scale=0.4]{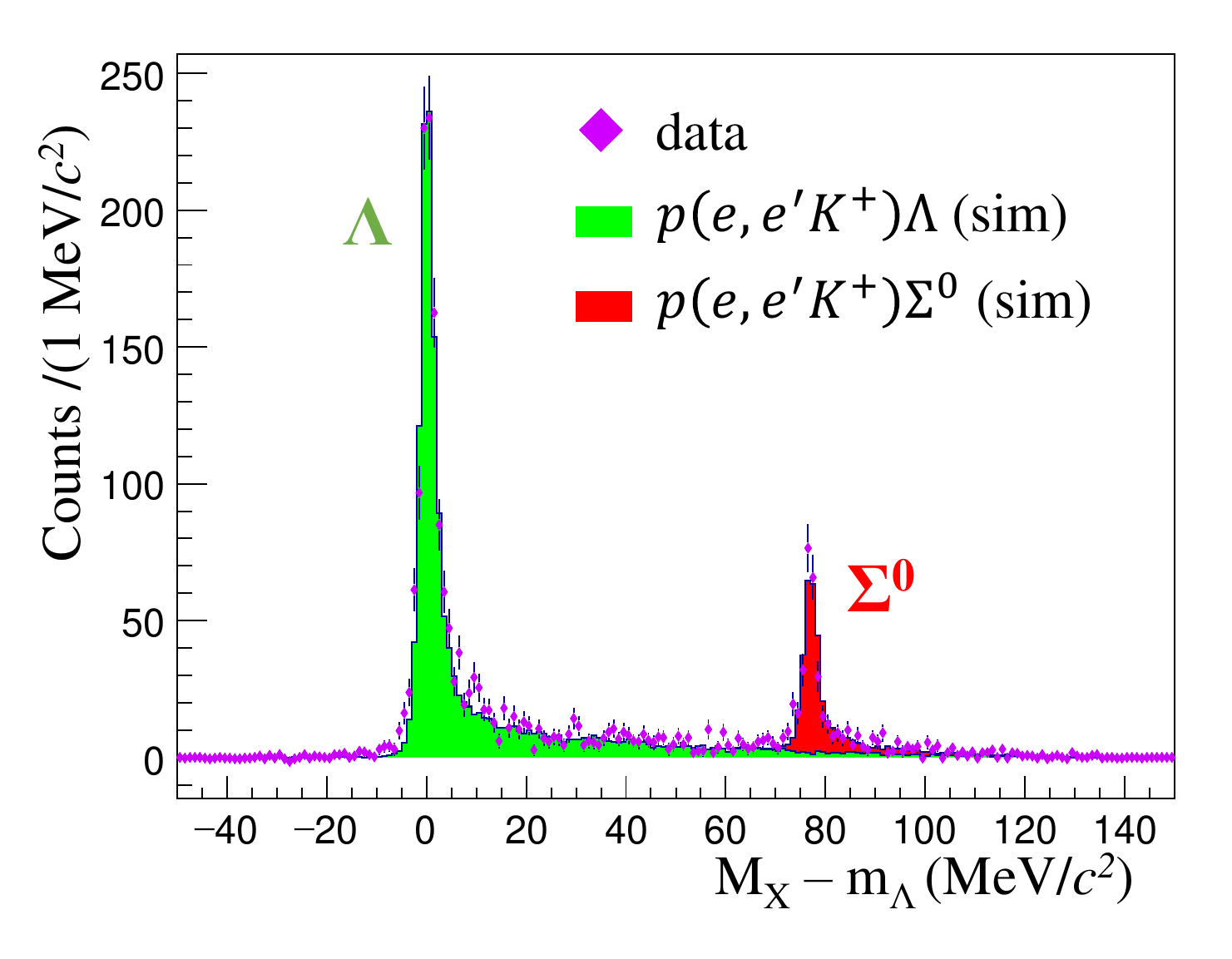}
\caption{\label{fig:simulation comparison} Missing-mass spectrum for the $p(e,e'K^+)\Lambda/\Sigma^0$ reaction with the momentum setting of ${\rm M_{calib.}}$. The data are shown by markers with statistical error bars, and the MC-simulation spectra are shown by histograms. }
\end{figure}

Elementary-reaction data of $\Lambda$ and $\Sigma^0$ from the hydrogen target were used for validating the MC simulation and the event generator.
Events were generated by the Geant4 MC simulation, and the missing-mass reconstruction was performed with the same analysis code as that for the real data analysis.
The input parameters of the simulation were position and angular resolutions at the focal plane based on the VDC resolution~\cite{Alcorn2004}.
The momentum vectors at the production points were calculated by the backward transfer matrices as shown in Eqs.~(\ref{TMtheta})--(\ref{TMp}).
Figure~\ref{fig:simulation comparison} shows the missing-mass spectrum obtained in the MC simulation compared with the experimental data.
The simulation agrees well with the experimental data.

\subsection{\label{sec:Energy loss correction}Energy loss correction}
Particles lose their energies in materials, and the energy losses need to be considered for the missing-mass reconstruction.
The measurable quantity in our experiment was the momentum.
Therefore, a correction for the momentum is a more practical option than the energy-loss correction in the analysis as was done in the past hypernuclear experiment~\cite{Gogami2018}.
The momentum-loss correction was applied event by event depending on $y'$ and $z_{\rm T}^{\rm mean}$ for $e'$ and $K^+$. 
The dependence on $y'$ and $z_{\rm T}^{\rm mean}$ came from the shape of the target cell~\cite{Santiesteban2019}.
On the other hand, the correction of a fixed value was applied to the incident electron beam.
The correction function was obtained from the MC simulation in which precise geometry was modelled as described in the previous section~(Sec.~\ref{sec:Monte Carlo simulation}).

\subsection{\label{Efficiencies} Efficiencies}
Efficiencies and correction factors needed for the cross-section calculation were evaluated and summarized in Tab.~\ref{tab:Efficiencies}.
The cross-section calculation with the acceptance of the HRS-R ($\Omega^{\rm{HRS\textrm{-}R}}$), the $K^+$ decay factor, and the $K^+$ absorption factor were applied event by event depending on the momenta of particles.
The other efficiencies and correction factors were applied as fixed values for all events.

\begin{table*}
\centering
\caption{Efficiencies and correction factors used for the cross section calculation in Eq.~(\ref{eqs:CS}).}
\label{tab:Efficiencies}
\begin{tabular}{lp{4cm}p{9cm}}
\hline
\textrm{Item} &
\begin{tabular}{l}
\textrm{Efficiency or} \\ \textrm{correction factor}
\end{tabular}
& \textrm{Regard}\\
\hline
$\epsilon_{\rm{track}}$& $0.981$ & Tracking efficiency 
estimated from the data analysis and a simple MC simulation.\\
$\epsilon_{\rm{decay}}$ & $0.15$ at $p^{\rm{cent.}}$ & 
$K^+$ survival ratio against its decay estimated by Geant4 MC simulation. \\
$\epsilon_{\rm{T}}$ & $0.986$  &
Survival ratio of the tritium gas against its decay with $^3\rm{H}\to ^3\rm{He}+e^-+\bar{\nu}_e $. \\
$1/\epsilon_{\rm{He}}$ & $0.97$ & 
Correction factor to correct the $^3{\rm{He}}$ contamination from the tritium decay. A ratio of Q.F. $\Lambda$ production from $^3{\rm{H}}$ to that of $^3{\rm{He}}$ was assumed to be the same as that of the $(e,e'p)$ reaction~\cite{Cruz2019}. \\
$\epsilon_{\rm{DAQ}}$ & $0.95$ & 
Efficiency of data acquisition system and trigger counters~\cite{CruzD}. \\
$\epsilon_{\rm{ctime}}$ & $0.96$ &
Efficiency for the real coincidence selection by the coincidence time (Fig.~\ref{fig:coin}). \\
$\epsilon_{\rm{absorp}}$ & $0.91$ at $p^{\rm{cent.}}$ & 
Survival ratio of $K^+$ against its absorption in materials due to the $K^+N$ interaction. It was estimated by the Geant4 MC simulation. \\
$\epsilon_{\rm{density}}$ & $0.901$ & 
Density-reduction effect of the gas due to the heat by beam irradiation~\cite{Santiesteban2019}. \\
$\epsilon_{\rm{vertex}}$ & $0.71$ & 
$z_{\rm{diff}}$ and $z_{\rm{mean}}$ cuts 
shown in Sec.~\ref{sec:Target}. \\
$\epsilon_{\rm{PID}}$ & $0.91$ & 
Survival ratio of signals after the particle identification by the gas and aerogel Cherenkov detectors. \\
1/$\epsilon_{\pi}$ & $0.98$ & 
Correction factor to correct the $\pi$ contamination. \\
\hline
\end{tabular}
\end{table*}

\section{\label{sec:Result} Results and discussions}
\subsection{\label{sec:Derivation of the differential cross section} Derivation of the differential cross section}

The differential cross section of the ($\gamma^{*},K^+$) reaction was obtained as follows,
\begin{eqnarray} \label{eqs:CS}
\overline{\left( \frac{d\sigma}{d\Omega_K}\right)} &=&
\frac{\int_{\textrm{HRS-R}} d\Omega_K \left(\frac{d\sigma}{d\Omega_K} \right)}
{{\int_{\textrm{HRS-R}}d\Omega_K}} \nonumber \\ 
&=& \frac{1}{N_T N_{\gamma^*}}
\sum_{i=1}^{N_{\rm{HYP}}}\frac{1}{\epsilon_i \Omega^{\textrm{HRS-R}}_{i}(p_K)},
\end{eqnarray}
where $N_T$ and $N_{\gamma^*}$ are the number of target nuclei and virtual photons, respectively.
The $\epsilon_i$ represents the product of the efficiencies and the correction factors shown in Tab.~\ref{tab:Efficiencies}.
$\Omega_i^{\textrm{HRS-R}}$ is the acceptance of HRS-R evaluated from the MC simulation.
The solid-angle acceptance was evaluated to be $\Omega^{\textrm{HRS-R}}\approx 6$~msr at the central momentum and has a dependence on the particle momentum.
Therefore, the acceptance correction was applied event by event depending on the particle momentum.
The number of virtual photons $N_{\gamma^*}$ can be calculated with $N_{\gamma^*}=\Gamma_{\rm{int}}N_e$ where
\begin{eqnarray}
\Gamma_{\rm{int}} = \iint \Gamma d\Omega_{e'} dE_{e'}.
\end{eqnarray}

\begin{figure}[bt]
\centering
\includegraphics[scale=0.4]{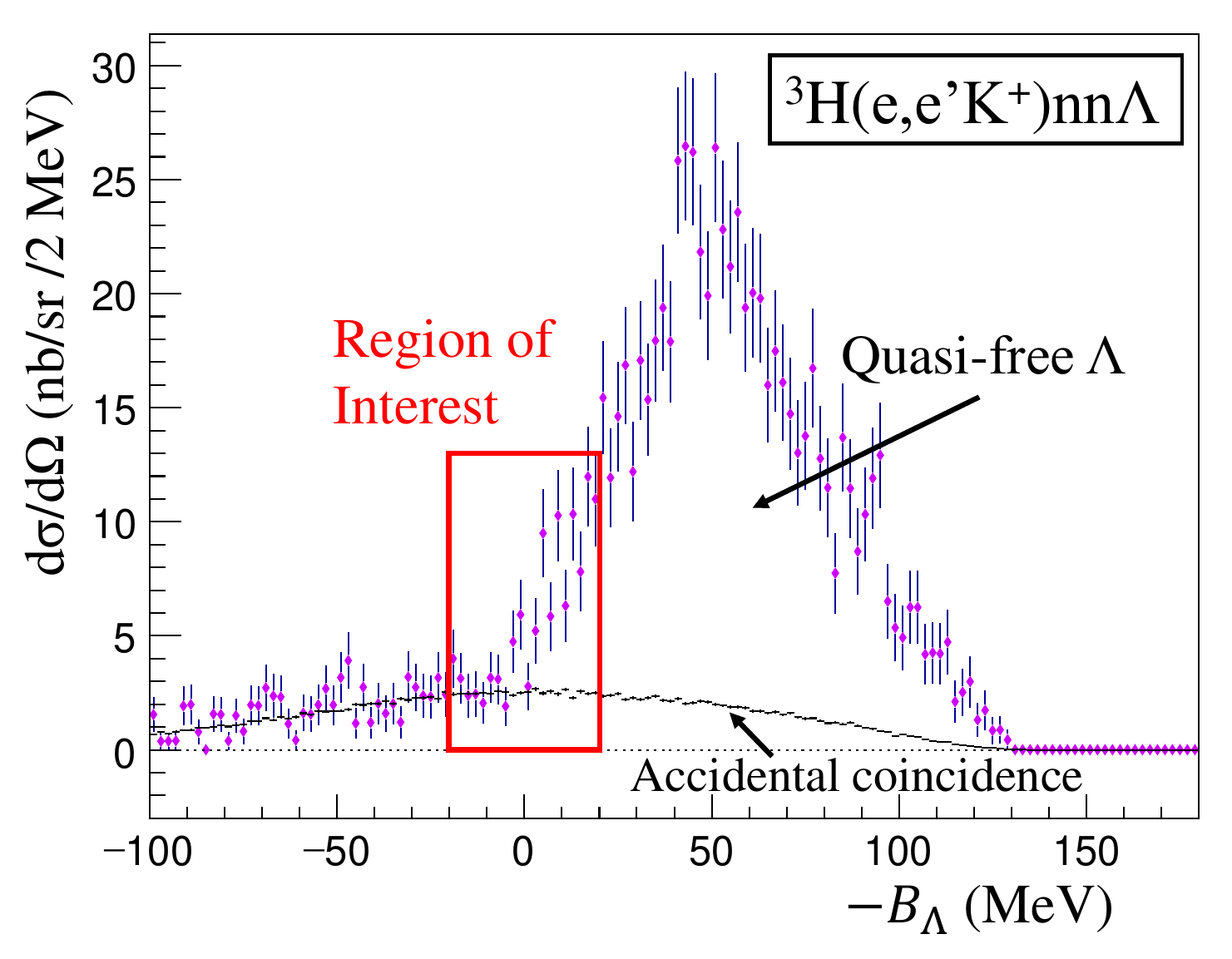}
\caption{\label{fig:cross section} Differential cross section of the reaction $^3\textrm{H}(e,e'K^+)\textrm{X}$ as a function of $-B_\Lambda$. A distribution of the accidental coincidence events was obtained via the mixed event analysis~\cite{Gogami2016Be,Gogami2016He}. The error bars shown in the figure are statistical only.}
\end{figure}

The differential cross section for the $^3{\rm{H}}(\gamma^*,K^+)$ reaction as a function of $-B_\Lambda$ is shown in Fig.~\ref{fig:cross section}.
A distribution of the accidental coincidence between $e'$ and $K^+$ is shown with an extremely small uncertainty.
The distribution of the accidental coincidence was evaluated by randomly combining $e'$ and $K^+$ in the analysis (mixed event analysis)~\cite{Gogami2016Be,Gogami2016He}.

\subsection{\label{sec::Energy Resolution}Energy Resolution}
The energy resolution was estimated from the MC simulation (Sec.~\ref{sec:Monte Carlo simulation}).
The simulation considered the energy straggling and multiple scattering effects in the materials as well as the optics of the spectrometers.
As a result, the expected resolutions were found to be $\sigma = 1.4\pm0.1$ and $1.5\pm0.2$~MeV for the $\Lambda$ and $nn\Lambda$ productions, respectively.
The errors on the expected resolutions mainly come from the variations of the possible magnetic field settings that could reproduce the data distributions of the momentum vectors at the production point and angle distributions of the particles at the focal plane.

\subsection{\label{sec::Systematic uncertainties}Systematic Uncertainties}
The systematic errors on the cross section and the binding energy were estimated, and
Tab.~\ref{tab:Systematic} shows the systematic uncertainties for the major efficiencies and the correction factors.
The largest contribution to the cross-section uncertainty comes from the acceptance of the spectrometer system.
The magnetic fields of the spectrometer magnets (QQDQ) in the MC simulation were changed and scanned so that the various particle distributions at the target and focal plane become consistent with those of the real data as shown in Sec.~\ref{sec:Monte Carlo simulation}.
In the scan, several magnetic-field settings that could reproduce the experimental data were found.
We derived the differential cross sections with the possible field settings that correspond to the possible acceptances,
and the difference was considered as the systematic error on the result.
The variations of the $\Omega^{\rm HRS\textrm{-}R}$ and $\Gamma_{\rm{int}}$ are $\pm 7.6\%$ and $\pm8.5\%$, respectively, due to the difference of the possible field settings.
The error on the acceptance shown above was evaluated over the whole acceptance.
The $\Omega^{\rm HRS\textrm{-}R}$ correction was applied event by event depending on the particle momentum in the present analysis.
Therefore, a resultant error due to the $\Omega^{\rm HRS\textrm{-}R}$ uncertainty is affected by the momentum distribution of data.

Major contributions to the systematic uncertainty on $B_\Lambda$ were obtained from (1) the calibration method and (2) the correction of momentum loss in materials particularly for the target cell.
The uncertainty that comes from the calibration method using the $\Lambda$ and $\Sigma^0$ productions was evaluated to be about $\pm0.1~\rm{MeV}$ in the previous hypernuclear experiment~\cite{Gogami2018}, which was performed following the same method.
We analysed the $p(e,e'K^+)\Lambda$ peak from the tritium gas target, in which a small percentage of hydrogen contamination took place.
It is noted that the hydrogen contamination made a broad peak at $-B_{\Lambda}\sim +45$~MeV in the spectrum of $^3\textrm{H}(e,e'K^+)\textrm{X}$ reaction, and the uncertainty related to the hydrogen contamination does not affect the present analysis of the differential cross section.
We applied a missing-mass correction so as to make the $\Lambda$ peak from the hydrogen contamination become consistent with the PDG mass~\cite{PDG2020}.
This correction yielded an uncertainty of $\Delta B_{\Lambda}^{\rm{A}} = \pm0.3$~MeV, which is from the statistical uncertainty of the $\Lambda$ peak arising from the hydrogen contamination in the tritium-gas cell.
The target-cell thickness needed to be assumed in the MC simulation to obtain the momentum-loss correction values.
However, the target cell was not uniform and had the sample-standard deviations of $7.6\%$ and $25\%$ for the cells of tritium and hydrogen gas targets, respectively.
The deviations of the cell thickness caused the uncertainty of the momentum-loss corrections for particles.
This effect was evaluated to be $\Delta B_{\Lambda}^{\rm{B}} = \pm 0.1$~MeV if the correction by the $\Lambda$ peak from the H contamination in the tritium-gas cell is performed with no uncertainties.
Therefore, the systematic error that comes from the momentum-loss correction was found to be $\sqrt{(\Delta B_{\Lambda}^{\rm{A}})^2 + (\Delta B_{\Lambda}^{\rm{B}})^2} = \pm 0.32$~MeV.

In total, the systematic error on $B_\Lambda$ was evaluated to be $\Delta B_{\Lambda}^{\rm{sys.}} = \pm0.4~\rm{MeV}$ for the present analysis.

\begin{table}
\centering
\caption{Relative errors for the efficiencies and the correction factors that contributed to the uncertainty on the cross-section calculation in addition to the acceptance uncertainties.}
\label{tab:Systematic}
\begin{tabular}{ll}
\hline
\textrm{Item} &
\textrm{Relative error} \\
\hline
$\Delta\epsilon_{\rm{track}}$ & $0.2\%$ \\
$\Delta\epsilon_{\rm{T}}$ & $0.3\%$ \\
$\Delta\epsilon_{\rm{He}}$ & $0.7\%$ \\
$\Delta\epsilon_{\rm{DAQ}}$ & $ 0.1\%$ \\
$\Delta\epsilon_{\rm{ctime}}$ & $0.5\%$  \\
$\Delta\epsilon_{\rm{density}}$ & $1.1\%$  \\
$\Delta\epsilon_{\rm{vertex}}$ & $7.8\%$  \\
$\Delta\epsilon_{\rm{PID}}$ & $6.9\%$  \\
$\Delta\epsilon_{\pi}$ & $1.8\%$ \\
$\Delta\Gamma_{\rm{int}}$ & $8.5\%$ \\
$\Delta N_e$ & $1.0\%$ \\
$\Delta N_T$ & $0.9\%$ \\
$\Delta N_{\rm{HYP}}$ & $6.8\%$\\
Single track selection & $1.3\%$ \\ 
\hline
Total & $15\%$ \\
\hline
\end{tabular}
\end{table}

\subsection{Upper Limit Analysis and Discussion}
The $nn\Lambda$ signal was searched for in a threshold region of $-B_{\Lambda}$ ranging from $-20$ to $20$~MeV.
We performed the spectral fits assuming the distributions for the quasi-free $\Lambda$ (QF), accidental coincidence, and signal to analyze the differential cross section.
The spectral fits were carried out by the unbinned maximum likelihood with RooFit toolkit~\cite{Verkerke2003}.
Background events of the QF start to rise at $-B_{\Lambda}=0$~MeV and monotonically increase as a function of $-B_\Lambda$ in the threshold region.
The QF distribution at the threshold region in particular is affected by the $\Lambda n$ FSI and  becomes complicated.
However, there are no theoretical predictions for the QF distribution with the FSI so far.
Therefore, in the present analysis, we assumed a linear function, which is the simplest assumption for the QF background.
A distribution function for the accidental coincidence was obtained by a fit with the 4th-order polynoamial function for the accidental coincidence distribution that was obtained by the mixed event analysis~(Sec.~\ref{sec:Derivation of the differential cross section}).

An experimental peak has a long tail mainly due to the external and internal radiations~\cite{Vanderhaeghen2000}, as shown in Fig.~\ref{fig:simulation comparison}.
A response function of the signal was obtained by the MC simulation.
In addition to the response function, which includes the experimental resolution and the tail component, the decay width $\Gamma$ of the $nn\Lambda$ state needs to be considered.
Here, we convoluted a Breit--Wigner~(BW) function with the decay width $\Gamma$ into the experimental response function to make a template function for the signal.
There is a slight difference between the data and MC simulation for the tail component in the $\Lambda$ production spectrum shown in Fig.~\ref{fig:simulation comparison}.
There may be the similar difference in the tail component for the $nn\Lambda$ production as well.
The error that may come from the difference in the tail shape was evaluated by a simple test.
In the test, the tail shape of MC simulation which was adjusted to the data for the $\Lambda$ production was adopted to the response function for the $nn\Lambda$ production when the spectral fit was performed.
As a result, the systematic error on the cross section originating from the tail-shape uncertanty was evaluated to be $\Delta N_{\rm HYP} = \pm6.8\%$.

\begin{figure*}[bt]
\centering
\includegraphics[scale=0.30]{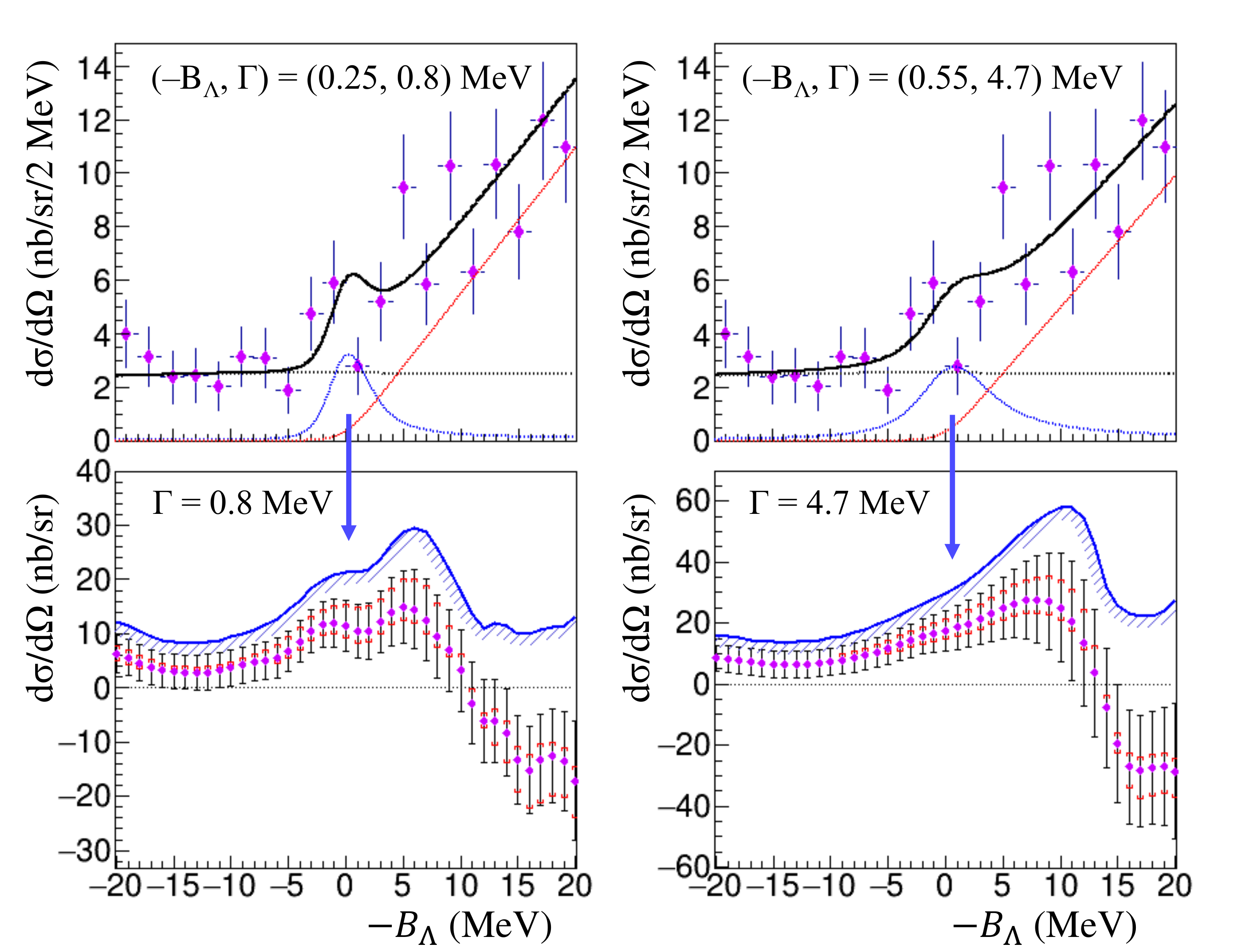}
\caption{\label{fig:UpperLimit} The differential cross section as a function of $-B_\Lambda$ (MeV). 
Spectral fits were done by assuming $(-B_\Lambda, \Gamma)=(0.25,0.8)$ and $(0.55,4.7)$ MeV respectively which are predictions adopted from Refs.~\cite{Kamada2016} and \cite{Belyaev2008}, respectively.
The figure in each panel shows the differential cross section of exceeded events over the assumed QF distribution as a function of an assumed peak center.}
\end{figure*}

The top part in Fig.~\ref{fig:UpperLimit} show the fitting result with assumptions of $(-B_\Lambda, \Gamma) = (0.25,0.8)$ and $(0.55,4.7)~\rm{MeV}$, which are theoretical predictions obtained from Refs.~\cite{Kamada2016} and \cite{Belyaev2008}, respectively.
The differential cross sections were obtained to be $11.2\pm4.8({\rm{stat.}})^{+4.1}_{-2.1}({\rm sys.})$ and
$18.1\pm6.8({\rm{stat.}})^{+4.2}_{-2.9}({\rm sys.})$~nb/sr for the assumptions of $(-B_\Lambda,\Gamma) = (0.25, 0.8)$ and $(0.55,4.7)$~MeV, respectively.
Given the decay widths, the differential cross sections as a function of the assumed peak position are shown in the bottom part of Fig.~\ref{fig:UpperLimit}.
The systematic errors are represented by the selection symbols.
There seems to be some excesses in the range from $-5$ to $5$~MeV for both assumptions of the decay width; one is narrow and the other is wide.
However, the excesses do not have a statistical significance of more than $3\sigma$.
It is noted that the differential cross section becomes negative for the region about $-B_{\Lambda}>10$~MeV.
This is because of a larger gradient of the linear function for QF, which was led by some events around $-B_{\Lambda}=0$~MeV.
The larger gradient of the linear function alone caused an overestimation at $-B_{\Lambda} > 10$--$12$~MeV, leading to a negative amplitude of the signal function.
In addition, a fraction of area being outside of the fit region ($-B_{\Lambda} > 20 $~MeV) for the tail component of the signal function increases as the assumed peak position becomes larger.
Therefore, the fit results for $-B_{\Lambda} > 10$--$12$~MeV have another systematic error in addition to the systematic errors that we considered in the present analysis.

\begin{figure}[bt]
\centering
\includegraphics[scale=0.45]{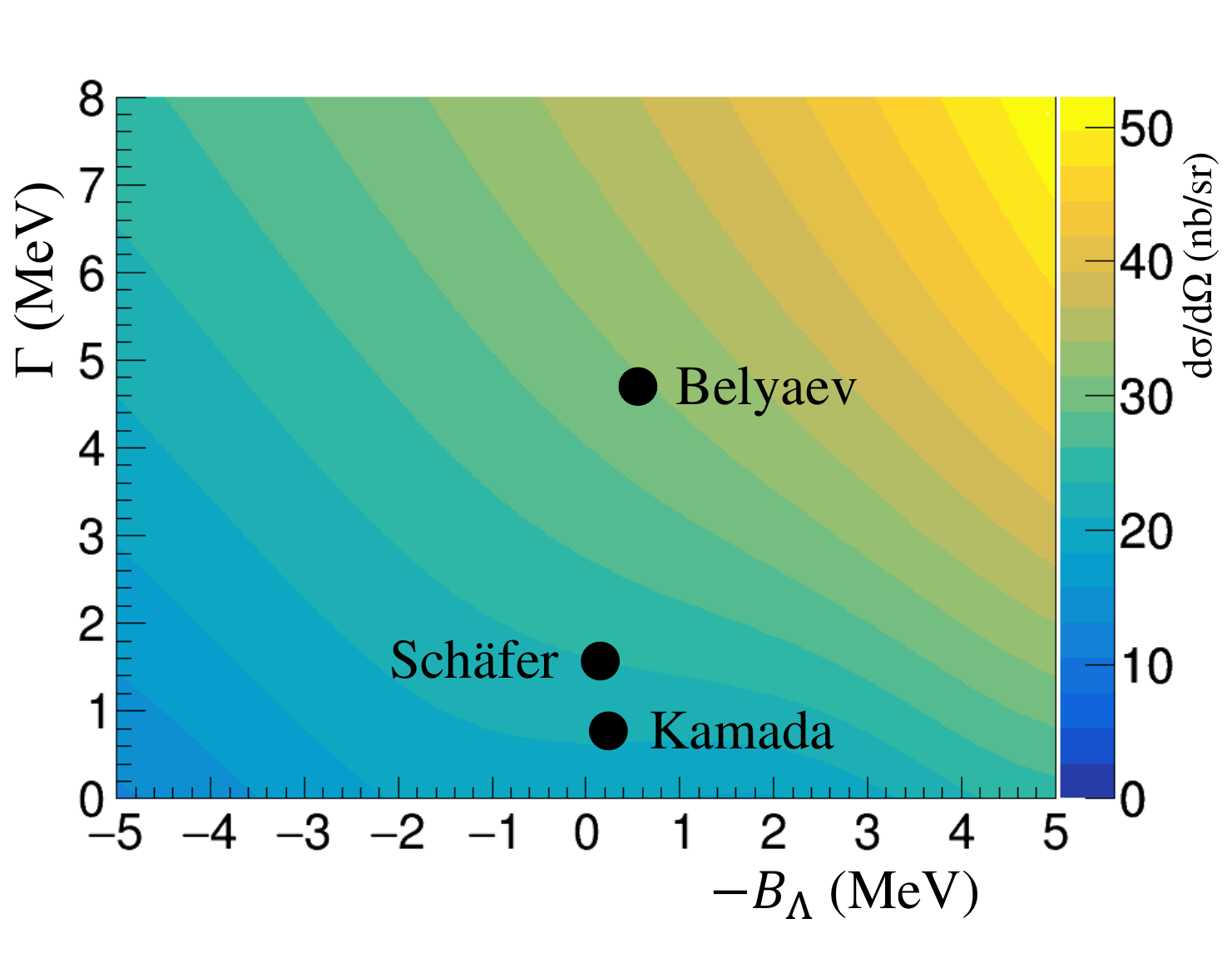}
\caption{\label{fig:ULScan} Two dimensional map of the upper limit of the differential cross section at the $90\%$ confidence level for $B_\Lambda$ and the decay width.
Theoretical predictions (Kamada, Belyaev and Sch$\textrm{\"{a}}$fer) shown in the figure were adopted from Refs.~\cite{Kamada2016}, \cite{Belyaev2008} and \cite{Martin2021}, respectively.}
\end{figure}

The $90\%$-confidence level (C.L.) of the differential cross-section upper limit was further evaluated.
The upper limit ($x_{\rm{U.L.}}^{\rm{stat.}}$) with the $90\%$ C.L. which takes into account the statistical error was calculated as follows:
\begin{eqnarray}
  \frac{\int_0^{x_{\rm{U.L.}}^{\rm{stat.}}} g(x)dx}{\int_0^\infty g(x)dx} = 0.90,
\end{eqnarray}
where $g(x)$ represents the Gaussian function.
The solid lines in the bottom part of Fig.~\ref{fig:UpperLimit} represent the final upper limits after the systematic uncertainties were also considered.
Here, the total systmatic uncertainty was simply added to the statistical error.
The $90\%$-C.L. upper limits were obtained to be $21$ and $31$ nb/sr for the assumptions of $(-B_{\Lambda},\Gamma) = (0.25,0.8)$ and $(0.55,4.7)$ MeV, respectively.
In addition to the above analyses in which the decay width is fixed,
two dimensional scans with the peak position and the decay width were performed, and the result is shown in Fig.~\ref{fig:ULScan}.

In the present analysis, in which the statistics are limited by event selection to avoid a large systematic error on the cross section,
no significant structures were observed with the simple assumptions of the QF shape.
This would be either because of the small cross section or due to the large decay width.
The possibility of nonexistence of either the resonant or bound state of $nn\Lambda$, which is suggested by Afnan {\it{et al.}} with the normal strength of the $\Lambda n$ interaction~\cite{Afnan2015}, cannot be excluded.
Theoretical predictions of the QF distribution with the $\Lambda n$ FSI are desired for further analysis to investigate the $nn\Lambda$ state.
In other words, the $\Lambda n$ interaction may be extracted by analyzing the QF distribution of the present data.

\section{\label{sec:Conclusion} Conclusion}
Missing-mass spectroscopy with the $^3{\rm{H}}(e,e'K^+)\textrm{X}$ reaction was performed at JLab Hall A to investigate the $nn\Lambda$ state.
The analysis focused on the measuring the differential cross section in the present article.
Therefore, only events that were detected in the acceptance of $|\delta p/p|< 4\%$ where the data was reproduced well from MC simulation were selected and used for the analysis.
The distribution of the quasi-free $\Lambda$ (QF) is not trivial, and no predictions exist so far.
Hence, we assumed a simple function (linear function) for the QF distribution to scan an excess above the QF.
As a result, no peaks with more than $3\sigma$ of the statistical significance were observed in the threshold region ($-20 \leq -B_{\Lambda} \leq 20$~MeV).
The $90\%$-C.L. upper limits were obtained to be $21$ and $31$ nb/sr for assumptions of $(-B_\Lambda,\Gamma)=(0.25,0.8)$ and $(0.55,4.7)$~MeV, respectively, which are the theoretically predicted energies and decay widths~\cite{Kamada2016,Belyaev2008}.
The present analysis provides valuable information to examine the existence of either $nn\Lambda$ bound or resonant state.
In addition, the cross-section result obtained here would give us a constraint for the $\Lambda n$ interaction by comparing the theoretical predictions with various interaction models.
Data analyses (i) to search for a peak from a count-base spectrum for which the larger statistics are available and (ii) to extract the $\Lambda n$ interaction from the QF shape are ongoing and will be discussed in further studies.

\section*{\label{sec:Acknowledgement} Acknowledgement}
We thank the JLab staff of the Division of Physics, Division of Accelerator, and the Division of Engineering for providing the support for conducting the experiment.
We acknowledge the outstanding contribution of the Jefferson Lab target group for the design and safe handling of the tritium target for the present experiment.
Additionally, we thank B.~F.~Gibson, E.~Hiyama, T.~Mart, T.~Motoba, K.~Miyagawa and M. Sch$\textrm{\"{a}}$fer for extensive discussions.
This work was supported by the U.S. Department of Energy (DOE) grant DE-AC05-06OR23177 under which Jefferson Science Associates, LLC, operates the Thomas Jefferson National Accelerator Facility.
The work of ANL group member is supported by DOE grant DE-AC02-06CH11357.
The Kent State University contribution is supported under Grant No.~PHY-1714809 from the U.S. National Science Foundation.
The hypernuclear program at JLab is supported by US-DOE grant DE-FG02-97ER41047.
This work was partially supported by the Grant-in-Aid for Scientific Research on Innovative Areas ``Toward new frontiers Encounter and synergy of state-of-the-art astronomical detectors and exotic quantum beams.''
This work was supported by JSPS KAKENHI Grants No.~18H05459, No.~18H05457, No.~18H01219, No.~17H01121, No.~19J22055, No.~18H01220.
This work was also supported by SPIRITS 2020 of Kyoto University,
and the Graduate Program on Physics for the Universe, Tohoku University (GP-PU).
\nocite{*}

\bibliography{Author_tex}


\bibliographystyle{ptephy}
%



\let\doi\relax


\end{document}